\documentclass[twocolumn,showpacs,preprintnumbers,amsmath,amssymb,floatfix]{revtex4}
\input epsf

\newcommand \rt {\right}
\newcommand \lt {\left}
\newcommand \be {\begin{equation}}
\newcommand \ee {\end{equation}}
\newcommand \ben {\begin{eqnarray}}
\newcommand \een {\end{eqnarray}}
\newcommand \nline {\nonumber \\}

\newcommand \free {${\cal F}$\ }
\newcommand \freem {{\cal F}}

\newcommand \freedm {{\it F}}
\newcommand \noise {{\cal D}}
\newcommand \pxpy[2] {\frac{\partial #1}{\partial #2}}
\newcommand \dxdy[2] {\frac{d #1}{d #2}}

\newcommand \dr {\delta\rho}
\newcommand \dre {\delta\rho_o}
\newcommand \ddr {\epsilon}
\newcommand \drq {\delta\hat{\rho}}

\newcommand \disp {\omega(\nabla^2)}
\newcommand \dispq {\hat{\omega}_q}
\newcommand \dispqs {\hat{\omega}_{q^*}}
\newcommand \dispqo {\hat{\omega}_0}

\begin{document}

\title{Modeling Elastic and Plastic Deformations in 
Non-Equilibrium Processing Using Phase Field Crystals.}

\author{K. R. Elder$^1$ and Martin Grant$^2$}

\affiliation{$^1$Department of Physics, Oakland University, Rochester, MI,
48309-4487}
\affiliation{$^2$Physics Department, Rutherford Building, 3600 rue
University, McGill University,
Montr\'eal, Qu\'ebec, Canada H3A 2T8.}

\date{\today}

\begin{abstract}

A continuum field theory approach is presented for modeling elastic 
and plastic deformation, free surfaces and multiple crystal orientations 
in non-equilibrium processing phenomena.  Many basic properties of 
the model are calculated analytically and numerical simulations are 
presented for a number of important applications including, 
epitaxial growth, material hardness, grain growth, reconstructive 
phase transitions and crack propagation. 

\end{abstract}

\pacs{05.70.Ln, 64.60.My, 64.60.Cn, 81.30.Hd}
\maketitle

\vskip1pc

\tableofcontents

\section{Introduction}

	Material properties are often controlled by the complex micro-structures that 
form during non-equilibrium processing.  In general terms the dynamics that occur 
during the processing are controlled by the nature and interaction of the 
topological defects that delineate the spatial patterns.  For example, in spinodal 
decomposition the topological defects are the surfaces that separate regions of 
different concentration.  The motion of these surfaces is mainly controlled by the 
local surface curvature and non-local interaction with other surfaces or boundaries 
through the diffusion of concentration.  In contrast, block-copolymer systems form lamellar or 
striped phases and the topological defects are dislocations in the striped lattice.
In this instance the defects interact through long-range elastic fields.  

	One method for modeling the topological defects is through the 
use of `phase fields'.  These can be thought of as physically 
relevant fields (such as concentration, density, magnetization, etc.), or simply 
as auxiliary fields constructed to produce the correct topological defect motion.  
In constructing phenomenological models it is often convenient to take the former point of 
view, since physical insight or empirical knowledge can be used to construct 
an appropriate mathematical description.  In this paper the construction of a phase field 
model \cite{ekhg02} to describe the dynamics of crystal growth that includes elastic and plastic 
deformations is described.  The model differs from other phase field approaches 
\cite{wk97,mg99,hmrg02} in that the model is constructed to produce phase fields that 
are periodic.  This is done by introducing a free energy that is a functional of 
the local time averaged density field, $\rho(\vec{r},t)$.  In this description 
the liquid state is represented by a uniform $\rho$ and the crystal state 
is described a $\rho$ that has the same periodic crystal symmetry as a given crystalline  
lattice.  This description of a crystal has been used in other contexts\cite{club,am78}, but 
not previously for describing material processing phenomena.  For simplicity this model will be 
referred to as the {\it Phase Field Crystal} model, or PFC model for short. 
	
	This approach exploits the fact that many properties of crystals are controlled 
by elasticity and symmetry.  As will be discussed in latter sections, any free energy 
functional that is minimized by a periodic field naturally includes elastic energy 
and symmetry properties of the periodic field.  Thus any property of a crystal that is 
determined by symmetry (e.g., relationship between elastic constants, number and type 
of dislocations, low-angle grain boundary energy, coincident site lattices, etc.) is 
also automatically incorporated in the PFC model.

	The purpose of this paper is to introduce and motivate this new 
modeling technique, discuss the basic properties of the model and to present 
several applications to technologically important non-equilibrium phenomena. 
In remainder of this section a brief introduction to phase field modeling 
techniques for uniform and periodic field is discussed and related 
to the study of generic liquid/crystal transitions.  In the following section 
a simplified PFC model is presented and the basic properties of the model are 
calculated analytically.  This includes calculation of the phase diagram, 
linear elastic constants and the vacancy diffusion constant.

	In Section (\ref{sec:app}) the PFC model is applied to a number of 
interesting phenomena including, the determination of grain boundary energies, 
liquid phase epitaxial growth and material hardness.  In each of these 
cases the phenomena are studied in some detail and the results are compared 
with standard theoretical results.  At the end of this section sample  
simulations of grain growth, crack propagation and a reconstructive phase 
transition are presented to illustrate the versatility of the PFC model.
Finally a summary of the results are presented in Section (\ref{sec:sum}).

\subsection{Uniform Fields and Elasticity}

	Many non-equilibrium phenomena that lead to dynamic spatial patterns can 
be described by fields that are relatively uniform in space, except near interfaces 
where a rapid change in the field occurs.  Classic examples include order/disorder 
transitions (where the field is the sublattice concentration), spinodal decomposition 
(where the field is concentration)\cite{GUNTON}, dendritic growth\cite{DENDRITES} 
and eutectics\cite{EUTECTICS}.  To a large extent the dynamics of these 
phenomena are controlled by the motion and interaction of the interfaces.  A great 
deal of work has gone into constructing and solving models that describe 
both the interfaces ('sharp interface models') and the fields ('phase field models').
Phase field models are constructed by considering symmetries and conservation 
laws and lead to a relatively small (or generic) set of sharp interface 
equations \cite{egpk01}. 

	To make matters concrete consider the case of spinodal decomposition in AlZn.  If 
a high temperature homogeneous mixture of Al and Zn atoms is quenched below the 
critical temperature small Al and Zn rich zones will form and coarsen in time.
The order parameter field that describes this phase transition is the concentration 
field.  To describe the phase transition a free energy is postulated (i.e., made up) 
by consideration of symmetries.  For spinodal decomposition the free energy is 
typically written as follows,
\be
\label{eq:basicF}
 {\cal F} = \int dV \left(f\left(\phi\right)+K|\vec{\nabla}\phi|^2/2\right),
\ee
where $f(\phi)$ is the bulk free energy and must contain two wells one 
for each phase (i.e., one for Al-rich zones and one for Zn-rich zones).  
The second term in Eq. (\ref{eq:basicF}) takes into account the fact that 
gradients in the concentration field are energetically unfavorable.  This is the 
term that leads to a surface tension (or energy/length) of domain walls that 
separate Al and Zn rich zones.  The dynamics are postulated to be dissipative and 
act such that an arbitrary initial condition evolves to a lower energy state. 
These general ideas lead to the well known equation of motion;
\be
\label{eq:basicEOM}
\frac{\partial \phi}{\partial t} = -\Gamma (-\nabla^2)^a \frac{\delta {\cal F}}{\delta \phi} + \eta_c,
\ee
where, $\Gamma$ is a phenomenological constant.  The Gaussian random variable, $\eta_c$, is 
chosen to recover the correct equilibrium fluctuation spectrum, has zero mean and 
two point correlation,
\be
\label{eq:basicNOISE}
\langle \eta_c(\vec{r},t)\eta_c(\vec{r}',t')=\Gamma k_BT(\nabla^2)^a\delta(\vec{r}-\vec{r}')\delta(t-t').
\ee
The variable $a$ is equal to one if $\phi$ is a conserved field, such as concentration, 
and is equal to zero if $\phi$ is a non-conserved field, such as sublattice concentration.   

A great deal of physics is contained in Eq. (\ref{eq:basicEOM}) and many papers have been 
devoted to the study of this equation.  While the reader is refereed to \cite{GUNTON} for details, 
the only salient points that will be made here is that 1) the gradient term and double well 
structure of $f(\phi)$ in Eq. (\ref{eq:basicF}) lead to a surface separating different 
phases and 2) the equation of motion of these interfaces is relatively independent of the 
form of $f(\phi)$.  For example it is well known \cite{egpk01} that if $\phi$ is non-conserved 
the normal velocity, $V_n$, of the interface is given by;
\be
\label{eq:ncI}
V_n = \kappa + A
\ee
where $\kappa$ is the local curvature of the interface and $A$ is directly proportional to the 
free energy difference between the two phases.  If $\phi$ is conserved then the motion of 
the interface is described by the following set of equations \cite{egpk01},
\ben
\label{eq:cI}
V_n=\hat{n}\cdot\vec{\nabla}[\mu(0^+)-\mu(0^-)], \nline
\mu(0) = d_o\kappa+\beta V_n \nline
\partial \mu/\partial t = D\nabla^2 \mu,
\een
where $\mu\equiv \delta F/\delta \phi$ is the chemical potential, $d_o$ is the capillary 
length, $\beta$ is the kinetic under-cooling coefficient, $\hat{n}$ is a unit vector 
perpendicular to the interface position, $D$ is the bulk diffusion constant and 
$0^+$ and $0^-$ are positions just ahead and behind the interface respectively.

	It turns out that Eqs. (\ref{eq:ncI}) and (\ref{eq:cI}) always emerge when the 
bulk free energy contains two wells and the local free energy increases when gradients 
in the order parameter field are present \cite{egpk01}.  In this sense Eqs. (\ref{eq:ncI}) 
and (\ref{eq:cI}) can be thought of as generic or universal features of systems 
that contain domain walls or surfaces.  As will be discussed in the 
next subsection a different set of generic features arise when the field prefers 
to be periodic in space.  Some generic features of periodic systems are that they 
naturally contain an elastic energy, are anisotropic and have defects that are 
topologically identical to those found in crystals.  A number 
of research groups have built these `periodic features` into phase field 
models describing uniform fields.  This approach has some appealing features, as 
one can consider mesoscopic length and time
scales.  But it can involve complicated continuum models.  For example, in Refs.\
\cite{mg99,hmrg02}, a continuum phase-field model was constructed to 
treat the motion of defects, as well as their interaction with moving
free surfaces.  Although such an approach gives explicit access to the stresses and
strains, including the Burger's vector via a ghost field, the interactions
between the nonuniform stresses and plasticity are 
complicated, since the former constitutes a free-boundary problem,
while the latter involves singular contributions to the strain, within the continuum
formulation.

\subsection{Periodic Systems}
\label{sec:periodic}

	In many physical systems periodic structures emerge.  Classic examples 
include  block-copolymers \cite{hacsshrc00}, Abrikosov vortex lattices in 
superconductors \cite{pcgbb98} and oil/water systems containing 
surfactants \cite{lggz91} and magnetic thin films.  In addition many 
convective instabilities \cite{ch93}, such as Rayleigh-B\'enard convection and a Margonoli 
instability lead to periodic structures (although it is not always possible to 
describe such systems using a free energy functional).  To construct a free 
energy functional for periodic systems it is important to make the somewhat trivial 
observation that unlike uniform systems, these systems are minimized by spatial 
structures that contain spatial gradients.  This simple observation implies 
that in a lowest order gradient expansion the coefficient of 
$|\vec{\nabla}\phi|^2$ in the free energy (see Eq. (\ref{eq:basicF})) is negative.  
By itself this term would lead to infinite gradients in $\phi$ so that the next 
order term in the gradient expansion must to be included (i.e., $|\nabla^2\phi|^2$).  
In addition to these two terms a bulk free energy with two wells is also needed, 
so that a generic free energy functional that gives rise to periodic structures 
can be written,
\ben
 {\cal F} &=& \int dV \left(
\frac{K}{\pi^2}\left[-|\vec{\nabla}\phi|^2
+\frac{a_o^2}{8\pi^2}|\nabla^2\phi|^2\right]
+f\left(\phi\right) \right) \nline
&=& \int dV \left(
\phi \frac{K}{\pi^2}\left[\nabla^2 
+\frac{a_o^2}{8\pi^2}\nabla^4\right]\phi
+f\left(\phi\right) \right), \nline
\een
where $K$ and $a_o$ are phenomenological constants.

	Insight into the influence of the gradient energy terms can be 
obtained by considering a solution for $\phi$ of the form $\phi=A\sin(2\pi x/a)$. 
For this particular functional form for $\phi$ the free energy becomes,
\ben
\frac{{\cal F}}{a} &=& 
KA^2 \left[-\frac{2}{a^2} + \frac{a_o^2}{a^4}+\cdots\right] 
+\frac{1}{a}\int dV f\left(\phi\right) \nline
&\approx& -\frac{KA^2}{a_o^2}+\frac{2KA^2}{a_o^4} \left(\Delta a\right)^2 
+\frac{1}{a}\int dV f\left(\phi\right), 
\een
where $\Delta a \equiv a-a_o$.  
At this level of simplification it can be seen that the free energy per unit length 
is minimized when $a=a_o$, 
or $a_o$ is the equilibrium periodicity of the system.  Perhaps more importantly 
it highlights the fact the energy can be written in a Hooke's law form 
(i.e., $E=E_o + (k \Delta a)^2$) that is so common in elastic phenomena.  Thus 
a generic feature of periodic systems is that for small perturbation (e.g., 
compressions  or expansions) away from the equilibrium they behave elastically.
This feature of periodic systems will be exploited to develop models for crystal 
systems in the next section.

\subsection{Liquid/Solid systems} 

	In a liquid/solid transition the obvious field of interest is the density field 
since it is significantly different in the liquid and solid phases.  More precisely the density is
relatively homogeneous in the liquid phase and spatially periodic (i.e., crystalline)  
in the solid phase.  The free energy functional can then be approximated as;  
\ben
 {\cal F} &=& \int d\vec{r}\left[H\left(\dr\right)\right] 
= \int d\vec{r}\left[f(\dr)+\frac{\dr}{2}\, G(\nabla^2) \dr\right]
\een
where $f$ and $G$ are to be determined and $\dr$ is the deviation of the 
density from the average density ($\bar{\rho}$).  
Under constant volume conditions $\dr$ is a conserved field, so that the 
dynamics are given by;
\be
\label{eq:eomrho}
\pxpy{\dr}{t} = \Gamma \nabla^2 \dxdy{{\cal F}}{\dr} + \eta, 
\ee
where $\eta$ is a Gaussian random variable with zero mean and two point 
correlation,
\be
\langle \eta(\vec{r},t)\eta(\vec{r}',t')\rangle = \Gamma k_b \nabla^2 \delta(\vec{r}-\vec{r}')
\delta(t-t').
\ee

	To determine the precise functional form of the operator $G(\nabla^2)$ it is 
useful to consider a simple liquid since $\dr$ is small and $f$ can 
be expanded to lowest order in $\dr$, i.e., 
\be
f_{liq} = f^{(0)} + f^{(1)}(\dr) + \frac{f^{(2)}}{2!}(\dr)^2 + \cdots
\ee
where $f^{(i)}\equiv (\partial^i f/\partial\dr^i)_{\dr=0}$.  In this limit Eq. (\ref{eq:eomrho}) 
takes the form
\be
\pxpy{\dr}{t} = -\Gamma \nabla^2\left[f^{(2)}+G(\nabla^2)\right] + \eta.
\ee
which can be easily solved to give,
\ben
\drq(\vec{q},t) &=& e^{-q^2\dispq\Gamma t}\drq(\vec{q},0) \nline
&& +  e^{-q^2\dispq\Gamma t} 
\int_0^t e^{q^2\dispq \Gamma t'} \eta(\vec{q},t'),
\een
where, $\vec{q}$ is the wavevector, $\dispq \equiv f^{(2)} + G(q^2)$,
$\drq$ is the Fourier transform of $\dr$, i.e., 
\be
\drq(\vec{q},t) \equiv \int d\vec{r} e^{i\vec{q}\cdot\vec{r}}\dr(\vec{r},t)/(2\pi)^d.
\ee
and $d$ is the dimension of space.
The structure factor, $S(q,t) \equiv \langle |\drq|^2 \rangle$, is then;
\be
S(q,t) = e^{-2q^2\dispq\Gamma t}S(q,0) 
+ \frac{k_bT}{\dispq}\left(1-e^{-q^2\dispq\Gamma t}\right).
\ee
In a liquid system the density is stable with respect to fluctuations which implies that 
$\dispq > 0$.  
The equilibrium liquid state structure factor, $S^{eq}_{liq}(q) = S(q,\infty)$, 
then becomes,
\be
S^{eq}_{liq}(q) = \frac{k_BT}{\dispq}. 
\ee
\begin{figure}[btp]
\epsfxsize=8cm \epsfysize=8cm
\epsfbox{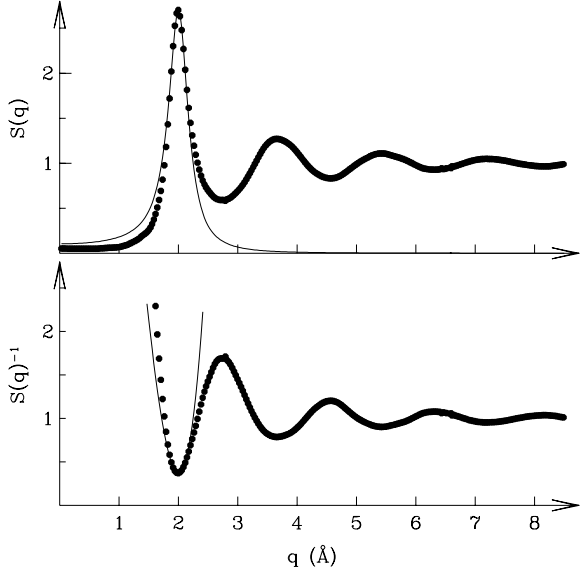}
\caption{
\label{fig:liqstAr}
The points correspond to an experimental liquid structure factor 
for $^{36}$Ar at 85$^o$K taken from J. Yarnell, M. Katz, R. Wenzel and S. Koenig, 
Phys. Rev. A, {\bf 7}, 2130 (1973) \cite{ykwk73}.  The line corresponds to a best fit 
to Eq. (\ref{eq:simpG}).}
\end{figure}

This simple calculation indicates the phase field method can model a liquid state if 
the function $\dispq$ is replaced with $k_BT/S^{eq}_{liq}(q)$, or 
\be
\label{eq:gtosk}
G(q) = k_B T/S^{eq}_{liq} - f^{(2)}. 
\ee  
A typical liquid state structure factor and the corresponding 
$\dispq$ is shown in Fig. (\ref{fig:liqstAr}).  Thus $G(\nabla^2)$ can 
be obtained for any pure material through Eq. (\ref{eq:gtosk}).

	In the solid state the density is unstable to the formation of a periodic 
structure (i.e., to forming a crystalline solid phase) and thus $\dispq$ must 
go negative for certain values of $q$.  This instability is taken into account 
by the temperature dependence of $f^{(2)}$, i.e., 
\be
f^{(2)} = a(T-T_m).
\ee
Thus when $T>T_m$, $w_q$ is positive and the density is uniform.  When $T<T_m$, 
$w_q$ is negative (for some values of $q$) and the density is unstable to 
the formation of a periodic structure.  To properly describe this state, 
higher order terms in $\dr$ must be included in the expansion of $f(\dr)$, 
since $\dr$ is no longer small.  Before discussing the properties of a specific choice 
for $f(\dr)$ it is worth pointing out some generic elastic features of such a model.

	As illustrated in the Section (\ref{sec:periodic}) a free energy that is 
minimized by a periodic structure has `elastic' properties.  The elastic constants of 
the system can be obtained by formally expanding around an equilibrium state in the strain 
tensor.  If the equilibrium state is defined to be $\dre(\vec{r})$ and 
the displacement field is $\vec{u}$, then $\dr$ can be written $\dr(\vec{r}) = 
\dre(\vec{r}+\vec{u})+\ddr$, where $\ddr$ will always be chosen to 
minimize the free energy.  Expanding to lowest order in the strain tensor gives,
\ben
 {\cal F} = {\cal F}_o +  \int d\vec{r} \Big(C_{ij,kl} \,u_{ij}\,u_{kl} +\cdots \Big)
\een
where $C_{ij,kl}$ is the elastic constant given by,
\be
\label{eq:elasticconstants}
C_{ij,kl} = \frac{1}{2!}\frac{\partial^2 H}{\partial u_{ij}u_{kl}}\Big|_{o}.
\ee
In Eq. (\ref{eq:elasticconstants}) the Einstein summation convection is used and 
$u_{ij}$ represents the usual components of the strain tensor, i.e., 
\be
u_{ij} \equiv \left(\frac{\partial u_i}{\partial u_j} + 
\frac{\partial u_j}{\partial u_i} 
+\frac{\partial u_l}{\partial u_i}\frac{\partial u_l}{\partial u_j}\right). 
\ee
and the subscript $o$ indicates that the derivative are evaluated at $\dr=\dre$ (i.e., 
$\dre=u_{ij}=0$).  While Eq. (\ref{eq:elasticconstants}) is somewhat formal and 
difficult to use for a specific model it does highlight several important features. 
Eq. (\ref{eq:elasticconstants}) shows that the elastic constants are simply related 
to the curvature of the free energy along given strain directions 
Perhaps more importantly Eq. (\ref{eq:elasticconstants}) shows that 
$C_{ij,kl}$ is proportional to $H$ which is a function of the equilibrium density 
field $\delta\rho_o$.   Thus if the free energy is written such that ${\cal F}$ is 
minimized by a $\dre$ that is cubic, tetragonal, hexagonal, etc., then $C_{ij,kl}$ 
will automatically contain the symmetry requirements of that particular system. 
In other words the elastic constants will always satisfy any symmetry requirement 
for a particular crystal symmetry since $C_{ij,kl}$ is directly proportional 
to a function that has the correct symmetry.  This also applies to the type or kind 
of defects or dislocations that can occur in any particular crystal system, since 
such deformation are determined by symmetry alone.

	In the next section a very simple model of a liquid/crystal transition 
will be presented and discussed in some detail.  This model is constructed by 
providing the simplest possible approximation for $f(\dr)$ that will lead to
towards a transition from a uniform density state (i.e., a liquid) to a 
periodic density state (i.e., a crystal).  

\section{Simple PFC model: Basic Properties}
\label{sec:basic}

	In this section perhaps the simplest possible periodic model of a 
liquid/crystal transition will be presented.  Several basic features of 
this model will be approximated analytically in the next few subsections. 
This includes calculations of the phase diagram, the elastic constants 
and the vacancy diffusion constant.

\subsection{Model}
	
	In the preceding section it was shown that a particular material can 
be modeled by incorporating the two point correlation function into the 
free energy through Eq. (\ref{eq:gtosk}).  It was also argued that the basic physical 
features of elasticity are naturally incorporated by any free energy  that 
is minimized by a spatially periodic function.  In this section the simplest 
possible free energy that produces periodic structures will be examined 
in detail.  This free energy can be constructed by fitting the following functional 
form for $G$,  
\be
\label{eq:simpG}
G(\nabla^2) = \lambda(q_o^2+\nabla^2)^2,
\ee 
to the first order peak in an experimental structure factor.  As an example 
such a fit is shown for argon in Fig.  (\ref{fig:liqstAr}).  At this level of 
simplification the minimal free energy functional is given by;
\be
\label{eq:shf}
\freem = \int d\vec{r} \left( \frac{\rho}{2} 
\left[a\Delta T +\lambda\left(q_o^2+\nabla^2\right)^2\right] \rho 
+ u \frac{\rho^4}{4}\right).
\ee
In principle other non-linear terms (such as $\rho^3$) can be included in the 
expansion but retaining only $\rho^4$ simplifies calculations.
The dynamics of $\rho$ are then described by the following equation,
\be
\pxpy{\rho}{t} = -\Gamma\nabla^2\mu+\eta.
= -\Gamma\nabla^2\dxdy{{\cal F}}{\rho}+\eta.
\ee
For convenience it useful to rewrite the free energy in dimensionless units, 
i.e., 
\ben
\vec{x} = \vec{r} q_o, \ \ 
\psi = \rho \sqrt{\frac{u}{\lambda q_o^4}}, \ \ 
r = \frac{a\Delta T}{\lambda q_o^4}, \ \ 
\tau=\Gamma \lambda q_o^6 t.
\een
In dimensionless units the free energy becomes, 
\be
\label{eq:dimfree}
F \equiv \frac{\freem}{\freem_o} = \int d\vec{x} 
\lt[\frac{\psi}{2}\disp\psi 
+ \frac{\psi^2}{4}\rt],
\ee
where $\freem_o \equiv \lambda^2q_o^{8-d}/u$ and 
\be
\label{eq:disp}
\disp=r+(1+\nabla^2)^2.
\ee 
The dimensionless equation of motion becomes,
\be
\label{eq:eom}
\pxpy{\psi}{t} = \nabla^2\lt(\disp\psi+\psi^3\rt)+\zeta,
\ee
where, $\langle \zeta(\vec{r}_1,t_1)\zeta(\vec{r}_2,\tau_2) \rangle = 
\noise \nabla^2\delta(\vec{r}_1-\vec{r}_2)\delta(\tau_1-\tau_2)$ and 
$\noise \equiv uk_BTq_o^{d-4}/\lambda^2$.

	Equations (\ref{eq:dimfree}), (\ref{eq:disp}) and (\ref{eq:eom}) describe a 
material with specific elastic properties.  In the next few sections the 
properties of this `material' will be discussed in detail.  As will be shown, some 
of the properties can be adjusted to match a given experimental 
system and others cannot be matched without changing the functional form of the 
free energy.  For example the periodicity (or lattice constant) can be adjusted 
since all lengths have been scaled with $q_o$.  The bulk modulus can be also be 
easily adjusted since the free energy has been scaled with $\lambda$, $u$ and 
$q_o$.  On the other hand this free energy will always produce a triangular lattice 
in two dimensions\cite{club,am78}.  To obtain a square lattice a different choice of the non-linear 
terms must be made.  This is the most difficult feature to vary as there are no 
systematic methods (known to the authors) for determining which functional 
form will produce which crystal symmetry.  Cubic symmetry can be obtained by 
replacing $\psi^4$ with $|\nabla\psi|^4$ \cite{sb97,gn03}.

	In the next few subsections the properties of this free energy and 
some minor extensions will be considered in one and two dimensions.
The three dimensional case will be discussed in a future paper.

\subsection{One dimension}
\label{sec:onedim}

	In one dimension the free energy described by Eq. (\ref{eq:dimfree}) is 
minimized by a periodic function for small values of $\psi_o$ and a constant for 
large values.  To determine the properties of the periodic state it is useful to 
make a one mode approximation, i.e., $\psi \approx A \sin(qx)+\psi_o$, which 
is valid in the small $r$ limit. Substitution of this function into Eq. (\ref{eq:dimfree}) 
gives,
\ben
\label{eq:onedf}
\frac{\freedm^p}{L} &=& \frac{q}{2\pi}\int_0^{2\pi/q} dx\,
\lt[\frac{\psi}{2}\omega(\partial_x^2)\psi 
+ \frac{\psi^2}{4}\rt]
\nline 
 &=& \frac{\psi_o^2}{2}\lt[\dispqo+\frac{3A^2}{2}+\frac{\psi_o^2}{2}\rt] 
+ \frac{A^2}{4}\lt[\dispq+\frac{3A^2}{8}\rt].
\een
where $\dispq$ is the fourier transform of $\disp$, i.e., 
$\dispq = r+(1-q^2)^2$.
Minimizing Eq. (\ref{eq:onedf}) with respect to $q$ gives the selected 
wavevector, $q^*=1$.  Minimizing \free with respect to $A$ gives,
$A^2 = -4(\dispqs/3 + \psi_o^2)$.  This solution is only meaningful if 
$A$ is real, since the density is a real 
field.  This implies that periodic solutions only exist when $r < -3\psi_o^2$, 
since $\dispqs = r$.  The minimum free energy density is then;
\be
\label{eq:fonemode}
F^p/L = -r^2/6
+\psi_o^2(1-r)/2-5\psi_o^4/4.
\ee
Equation (\ref{eq:fonemode}) represents the free energy density of a periodic 
solution in the one mode approximation.
To determine the phase diagram this energy must be compared to that for a 
constant state (i.e., the state for which $\psi^c = \psi_o$) which is,
\be
\label{eq:fconst}
F^c/L = \dispqo\psi_o^2/2+\psi_o^4/4.
\ee
\begin{figure}[btp]
\epsfxsize=8cm \epsfysize=8cm
\epsfbox{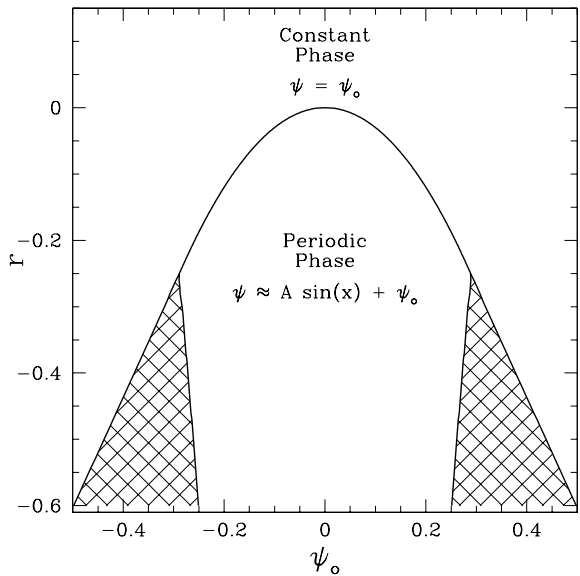}
\caption{One dimensional phase diagram in the one-mode approximation.
The solid line is the boundary separating constant (i.e., liquid) and 
periodic (i.e., crystal) phases.  The hatched section of the plot 
corresponds to regions of liquid/crystal coexistence.}  
\label{fig:phasdia}
\end{figure}

	To obtain the equilibrium states the Maxwell equal area construction 
rule must be satisfied, i.e., 
\be
\int_{\psi_1}^{\psi_2} d\psi \lt[\mu\lt(\psi_o\rt) - \mu_o\rt] = 0,
\ee
where $\psi_1$ is a solution of $\mu_p = \mu_o$, $\psi_2$ is a solution of 
$\mu_c = \mu_o$ and $\mu(\psi)=\mu_p$ ($=\mu_c$) if $\freem_p < \freem_c$ 
($\freem_p > \freem_c$) and $\mu = \partial \freem/\partial \psi_o$.
Using these conditions it is straightforward to show that for $r > -1/4$ 
a periodic state is selected for $|\psi_o| < \sqrt{-r/3}$ and a 
constant state is selected when $|\psi_o|> \sqrt{-r/3}$.
For $r < -1/4$, there can exist a coexistence of periodic and constant states.  
If the constant and periodic states are considered to be a liquid and crystal respectively 
then this simple free energy allows for the coexistence of a liquid and crystal, which 
implies a free surface.  The entire phase diagram is shown in Fig. (\ref{fig:phasdia}).

	It is also relatively easy to calculate the elastic energy in the one mode 
approximation.  
If $a\equiv 2\pi/q$ is defined as the one-dimensional lattice parameter,  then 
the \free can be written, 
\be
\freedm^p/L = \freedm^p_{min}/L +  K u ^2/2 + {\cal O} u^3 \cdots
\ee
where $u \equiv (a-a_o)/a_o$ is the strain and $K$ is the bulk modulus 
and is equal to;
\be
K = -
\lt(\psi_o^2+\dispqs/3\rt)\dxdy{^2\dispq}{q^2}\Big|_{q=q^*},
\ee
or for the particular dispersion relationship used here, 
$K= -8(r+3\psi_o^2)/3$.  The existence of such a Hooke's law relationships 
is automatic when a periodic state is selected since $F$ always increases when 
the wavelength deviates from the equilibrium wavelength. 

\subsection{Two dimensions}

\subsubsection{Phase Diagram}
\label{sec:twodimpd}

\begin{figure}[btp]
\epsfxsize=8cm \epsfysize=8cm
\epsfbox{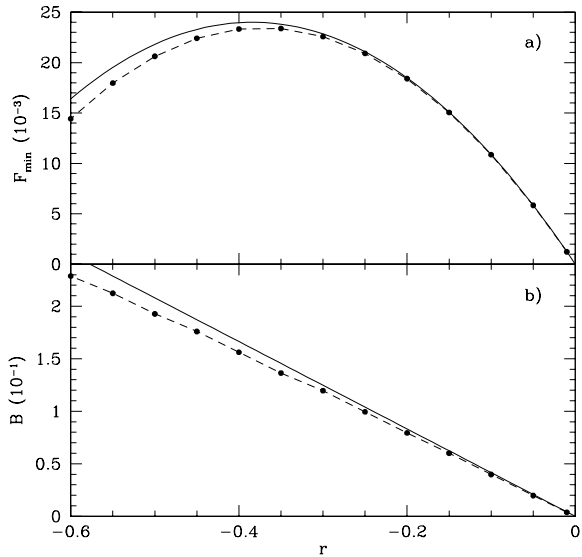}
\caption{In Fig. (a) the minimum of the free energy is plotted as a function of 
$r$ for $\psi_o = \sqrt{-r}/2$.  The solid line is Eq. (\ref{eq:freet}) and 
the points are from numerical simulations.  In Fig. (b) the bulk modulus is plotted 
as a function of $r$ for $\psi_o = \sqrt{-r}/2$.  The solid line is an analytic 
calculation ($(q_tA_t)^2$) and the points are from numerical simulations.}
\label{fig:cmpEE}
\end{figure}

In two dimensions $F$ is minimized by three distinct solutions 
for $\psi$.  These solutions are periodic in either zero dimensions (i.e., a constant),
one dimension (i.e., stripes) or two dimensions (i.e., triangular distributions of drops 
or `particles`).  The free energy density for the constant and stripe solutions 
are identical to the periodic and constant solution discussed in the 
preceding section.  The two dimensional solution can be written in the 
general form,  
\be
\label{eq:prdfn}
\psi(\vec{r}) = \sum_{n,m} a_{n,m} e^{i\vec{G}\cdot\vec{r}}+\psi_o,
\ee  
where, $\vec{G}\equiv n\vec{b}_1+m\vec{b}_2$ and the vectors $\vec{b}_1$ and $\vec{b}_2$ are 
reciprocal lattice vectors.  For a triangular lattice the reciprocal lattice vectors 
can be written,
\ben
\vec{b}_1&=&\frac{2\pi}{a\sqrt{3}/2}\lt(\sqrt{3}/2 \hat{x} +\hat{y}/2\rt) \nline
\vec{b}_2&=&\frac{2\pi}{a\sqrt{3}/2}\hat{y}
\een
where $a$ is the distance between nearest neighbor local maxima of $\psi$ (which 
corresponds to the atomic positions).  In analogy with the one-dimensional calculations 
presented (see Sec. (\ref{sec:onedim})) a one mode approximation will be made to evaluate 
the phase diagram and elastic constants.  In a two dimensional triangular system a 
one mode approximation corresponds to  retaining all fourier components that have the 
same length.  More precisely the lowest order harmonics consist of all $(n,m)$ pairs 
such that the vector $\vec{G}$ has length $2\pi/(a\sqrt{3}/2)$. This set of vectors 
includes $(n,m)=(\pm 1,0)$, $(0,\pm 1)$, $(1,-1)$ and $(-1,1)$. 
Furthermore since $\psi$ is a real function the fourier coefficients must satisfy 
the following relationship, $a_{n,m}=a_{-n,m}=a_{n,-m}$.  In addition, by symmetry,
$a_{\pm 1,0} = a_{0,\pm 1} = a_{1,-1} = a_{-1,1}$.  Taking these considerations into 
account it is easy to show that in the lowest order harmonic expansion for a 
triangular solution for $\psi$ can be represented by;
\ben
\label{eq:tri}
\psi_t &=& A_t \Big[\cos\lt(q_tx\rt)\cos\lt(q_ty/\sqrt{3}\rt) \nline
&&-\cos\lt(2q_ty/\sqrt{3}\rt)/2\Big] + \psi_o,
\een  
where $A_t$ is an unknown constant and $q_t = 2\pi/a$.
Substituting Eq. (\ref{eq:tri}) into Eq. (\ref{eq:shf}), 
minimizing with respect to $A_t$ and $q_t$ gives,
\ben
\label{eq:freet}
\frac{\freedm^t}{S} &\equiv& \int_0^{\frac{a}{2}} \frac{dx}{a/2} \int_0^{\frac{\sqrt{3}}{2}a} 
\frac{dy}{a\sqrt{3}/2} 
\lt[\frac{\psi}{2}\omega(\nabla^2)\psi +\frac{\psi^4}{4}\rt] \nline 
&=& -\frac{1}{10}\lt(r^2+\frac{13}{50}\psi_o^4\rt)
+ \frac{\psi_o^2}{2}\lt(1+\frac{7}{25}r\rt) \nline
&&+ \frac{4\psi_o}{25}\sqrt{-15r-36\psi_o^2}\lt(\frac{4\psi_o^2}{5}+\frac{r}{3}\rt),
\een
where,
\be
A_t = \frac{4}{5}\lt(\psi_o+\frac{1}{3}\sqrt{-15r-36\psi_o^2}\rt)
\ee
$q_t = \sqrt{3}/2$, and $S$ is a unit area.  The accuracy  
of this one mode approximation was tested by comparison with a 
direct numerical calculation for a range of $r$'s, using `Method I` 
as described in Appendix (\ref{app:nummet}).  The time step ($\Delta t$) and 
grid size ($\Delta x$) were $0.0075$ and $\pi/4$ respectively and a periodic   
grid of a maximum size of $512\Delta x \times 512 \Delta x$ \cite{periodic} was
used.  A comparison of the analytic and numerical solutions are shown in 
Fig. (\ref{fig:cmpEE}) for a variety of values of $r$ ($\psi_o$ was set to be $\sqrt{-r}/2$).  
The approximate solution is quite close to 
the numerical one and becomes exact in the limit $r\rightarrow 0$.  
The analytic results can in principle be systematically improved by including 
more harmonics in the expansion.  

\begin{figure}[btp]
\epsfxsize=8cm \epsfysize=8cm
\epsfbox{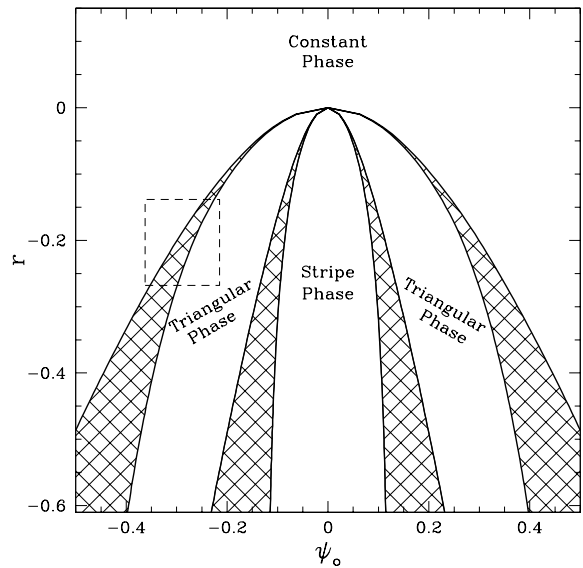}
\caption{Two dimensional phase diagram as calculated in one mode approximation. Hatched 
areas in the figure correspond to coexistence regions.  The small region enclosed by 
a dashed box is superimposed on the argon phase diagram in Fig. (\ref{fig:Arphd}).  In 
this manner the parameter of the free energy functional can be chosen to reproduce 
certain the relevant aspect of a liquid/crystal phase transition.}
\label{fig:phasdia2d}
\end{figure}

\begin{figure}[btp]
\epsfxsize=8cm \epsfysize=8cm
\epsfbox{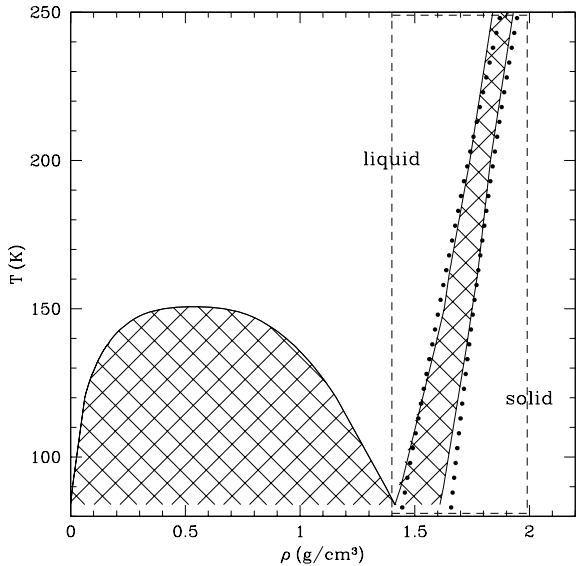}
\caption{The phase diagram of argon.  The hatched regions correspond to 
the coexistence regions.  The points are from the PFC model.
phase field model.}  
\label{fig:Arphd}
\end{figure}

To determine the phase diagram in two dimensions the free energy of the triangular 
state (i.e., Eq. (\ref{eq:freet}) must be compared with the free energy of a striped state 
(i.e., Eq. (\ref{eq:fonemode})) and a constant state (i.e., Eq. (\ref{eq:fconst})).  In 
addition since $\psi$ is a conserved field Maxwell's equal area construction must be used 
to determine the coexistence regions.  The phase diagram arising from these 
calculations is shown in Fig. (\ref{fig:phasdia2d}).
While this figure does not look like a typical liquid/solid phase diagram in the density-temperature 
plane it can be superimposed onto a portion of an experimental phase diagram.  
As an example the PFC phase diagram is superimposed onto the argon phase diagram in 
Fig. (\ref{fig:Arphd}).

\subsubsection{Elastic Energy}
	
	The elastic properties of the two dimensional triangular state 
can be obtained by considering the energy costs for deforming the equilibrium 
state.  The free energy density associated with bulk, shear and deviatoric 
deformations can be calculated by considering modified forms of Eq. (\ref{eq:tri}), 
i.e., $\psi_t(x/(1+\zeta),y/(1+\zeta))$ (bulk), $\psi_t(x+\zeta y,y)$ (shear) 
and $\psi_t(x(1+\zeta),y(1-\zeta)$ (deviatoric).  In such calculations $\zeta$ 
represents the dimensionless deformation, $q_t = \sqrt{3}/2$ and $A_t$ is obtained 
by minimizing $\it F$.  The results of these calculations are given below; 
\ben
\freedm_{bulk}/A &=& \freedm^t_{min} + \alpha\ \zeta^2 + \cdots \nline
\freedm_{shear}/A &=& \freedm^t_{min} + \alpha/8\ \zeta^2 + \cdots  \nline
\freedm_{deviatoric}/A &=& \freedm^t_{min} + \alpha/2\ \zeta^2 + \cdots \nline
\een
where
\be
\alpha = 4\lt(3\psi_o+\sqrt{-15r-36\psi_o^2}\rt)^2/75.
\ee
These results can be used to determine the elastic constants by noting 
that for a two dimensional system \cite{club,seitz},
\ben
\freedm_{bulk} &=& \freedm^t_{min} + [C_{11}+C_{12}]\,\zeta^2 + \cdots \nline
\freedm_{shear} &=& \freedm^t_{min} + [C_{44}/2]\,\zeta^2 + \cdots \nline
\freedm_{deviatoric} &=& \freedm^t_{min} + [C_{11}-C_{12}]\,\zeta^2 + \cdots .
\een
The elastic constants are then
\be
C_{11}/3=\, C_{12}= \,C_{44} = \alpha/4
\ee
These results are consistent with 
the symmetries of a two dimensional triangular system, 
i.e.,  $C_{11} = C_{12}+2\,C_{44}$. 
In two dimensions this implies a bulk modulus of $B = \alpha/2 $, 
a shear modules of $\mu = \alpha/4 $, a Poisson's ratio of 
$\sigma = 1/3$, and a two dimensional (i.e., $Y_2 = 4B\mu/(B+\mu)$) 
Young's modulus of $Y_2 = 2\alpha/3$.  Numerical simulations were 
conducted (using the parameters and numerical technique discussed in 
the previous section) to test the validity of these approximations for the 
bulk modulus.  The results, shown in Fig. (\ref{fig:cmpEE}), indicate 
that the approximation is quite good in the small $r$ limit.
  
These calculations highlight the strengths and limitations of the simplistic 
model described by Eq. (\ref{eq:shf}).  On the positive side the model 
contains all the expected elastic properties (with the correct 
symmetries) and the elastic constants can be approximated 
analytically within a one mode analysis.  One the negative side, the model as 
written, can only describe a system where $C_{11} = 3\, C_{12}$.  
Thus parameters in the free energy can be chosen to produce any 
$C_{11}$, but $C_{12}$ cannot be varied independently.  To obtain more 
flexibility a term $\psi^3$ could be added to the free energy.

\subsubsection{Dynamics}
\label{sec:dynamics}

\begin{figure}[btp]
\epsfbox{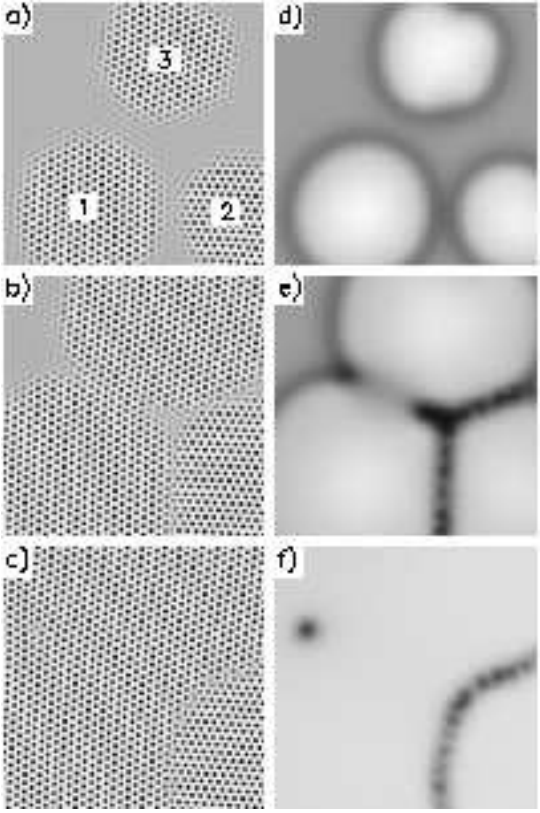}
\caption{Heterogeneous nucleation of a three crystallites in a supercooled liquid.  
The grey scale in a), b) and c) corresponds to the density field, $\psi$ and 
in d), e) and f) to the smoothed local free energy.  The configurations are taken 
at times $t=300, 525$ and $3975$ for a)+d), b)+e) and c)+f) respectively.
(Note only a portion of the simulation is shown here)}
\label{fig:multi}
\end{figure}

	The relatively simple dynamical equation for $\psi$ 
(i.e., Eq. (\ref{eq:eom})) can describe a large number of physical 
phenomena depending on the initial conditions and boundary conditions. 
To illustrate this versatility it is useful to consider the growth a crystalline phase 
from a supercooled liquid, since this phenomena simultaneously involves the motion of 
liquid/crystal interfaces and grain boundaries separating crystals of different orientations.
Numerical simulations were conducted using the `Method I` as described in 
Appendix (\ref{app:metI}).  The parameters for these simulations were,
$(r,\psi_o,D,\Delta x,\Delta t)=(-1/4,0.285,10^{-9},\pi/4,0.0075)$ on 
a system of size $512\Delta x \times 512 \Delta x$ with periodic boundary 
conditions.  The initial condition consisted of large random gaussian fluctuations 
(amplitude 0.1) covering $(10x10)$ grid points in three locations in the simulation cell.   As shown 
in Fig. (\ref{fig:multi}) the initial state evolves into three crystallites each with 
a different orientation and a well defined liquid/crystal interface.
The excess energy of the liquid/crystal interfaces is highlighted in 
Fig. (\ref{fig:multi}d) where the local free energy density is plotted.  

As time evolves the crystallites impinge and form grain boundaries.  As can be seen 
in Fig. (\ref{fig:multi}) the nature of the grain boundary between 
grains (1) and (3) is significantly different from the boundary between grains (2) and 
grain (1) (or (3)).  The reason is that the orientation of grains (1) and 
(3) is quite close but significantly different from (2).  The low angle grain boundary 
consists of dislocations separated by large distances, while the high angle grain 
boundary consists of many dislocations piled together.  A more detailed 
discussion of the grain boundaries will be given in Section (\ref{sec:grbo}). 
Even this small sample simulation illustrates the flexibility and power of the PFC 
technique.  This simulation incorporates, the heterogeneous nucleation of crystallites, 
crystallites with triangular symmetry and elastic constants,  crystallites of multi 
orientations, the motion of liquid/crystal interfaces and the creation and motion 
of grain boundaries.  While all these features are incorporated in standard microscopic 
simulations (e.g., molecular dynamics) the time scales of these simulations are much 
longer than could be achieved using microscopic models.

\begin{figure}[btp]
\epsfxsize=8cm \epsfysize=8cm
\epsfbox{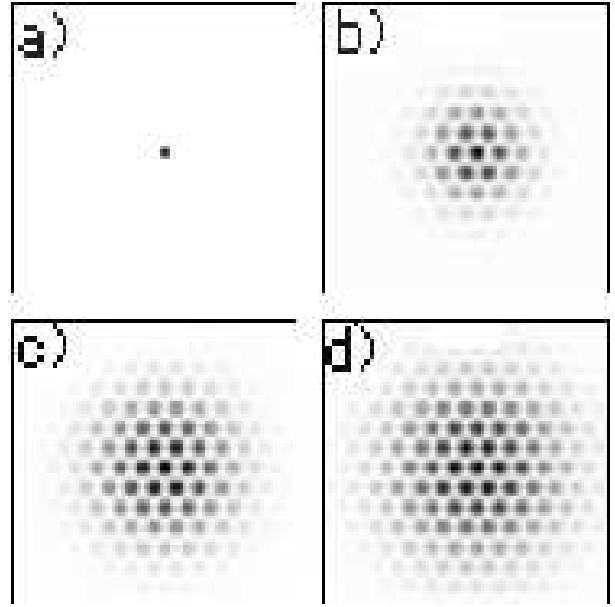}
\caption{Vacancy diffusion times.  In this figure the grey scale 
is proportional to the $\psi(\vec{r},t)-\psi_{eq}$.  The times 
shown are a) t=0, b) t=50, c) t=100 and d) t=150.}
\label{fig:diff1}
\end{figure}

One fundamental time scale in the PFC model is the diffusion time.  To envision 
mass diffusion in the PFC model it convenient to consider a perfect 
equilibrium ($\psi^t$) configuration with one `particle` missing.  At the atomic level 
this would correspond to a vacancy in the lattice.  Phonon vibrations would occasionally 
cause neighboring atoms to hop into the vacancy and eventually the vacancy 
would diffuse throughout the lattice.  In the PFC model the time scales associated 
with lattice vibrations are effectively integrated out and all that is left 
is long time mass diffusion.  In this instance the density at the missing spot 
will gradually increase as the density at neighboring sites slowly decreases.  
Numerical simulations of this process are shown in Fig. (\ref{fig:diff1}) using 
Method I (see App. (\ref{app:metI})) with the parameters $(r,\psi_o,\noise,\Delta x,
\Delta t)=(1/4,1/4,0,\pi/4,0.0075)$.  To highlight diffusion of the vacancy,
the difference between $\psi(\vec{r},t)$ and a perfect equilibrium 
state ($\psi^t$) is plotted in this Fig (\ref{fig:diff1}).

	The diffusion constant in this system can be obtained by a simple linear stability 
analysis, or Bloch-Floquet analysis, around an equilibrium state.  To begin the analysis the 
equation of motion for $\psi$ is linearized around $\psi^t$, i.e., 
$\psi = \psi^t(\vec{r})+\delta\psi(\vec{r},t)$.  To first order in $\delta\psi$ Eq. (\ref{eq:eom}) 
becomes, 
\be
\label{eq:fb}
\pxpy{\delta\psi}{t}= \nabla^2\lt[\lt(\omega
+3\lt(\psi_o^2+2\psi_o\phi_t+\phi_t^2\rt)\rt)\delta\psi\rt],
\ee
where $\phi_t = \psi_t-\psi_o$ (see Eq. (\ref{eq:prdfn})).  
The perturbation, $\delta\psi$, is then expanded as follows, 
\be
\label{eq:flprt}
\delta\psi = \sum_{n,m} b_{n,m}(t) e^{iq_t(n x + (n+2m)y/\sqrt{3})+i\vec{Q}\cdot \vec{r}}.
\ee
Substituting Eg. (\ref{eq:flprt}) into Eq. (\ref{eq:fb}) gives;
\ben
\label{eq:evals}
\pxpy{b_{i,j}}{t} &=& -k_{i,j}^2 
\Big(\lt(3\psi_o^2+\hat{\omega}\rt)b_{i,j} 
+ 6 \psi_o\sum_{n,m} a_{n,m}b_{i-n,j-m} \nline
&&+3 \sum_{n,m,l,p} a_{n,m}a_{l,p} b_{i-n-l,j-m-p}\Big)
\een 
where, $\hat{\omega} \equiv r+(1-k_{i,j}^2)^2$
and $ k_{i,j}^2 \equiv (i q_t+Q)^2 + q_t^2(i+2j)^2/3$.

\begin{figure}[btp]
\epsfxsize=8cm \epsfysize=8cm
\epsfbox{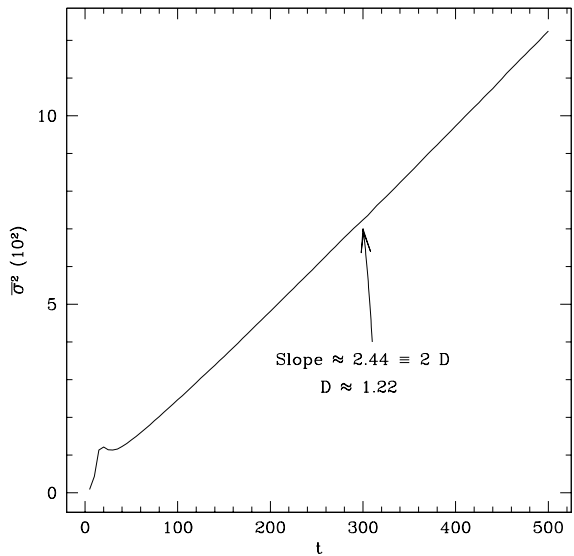}
\caption{Vacancy diffusion.  In this figure the average of the 
standard deviation in the x and y directions is plotted as a function 
of time.}
\label{fig:diff3}
\end{figure}

	To solve Eq. (\ref{eq:evals}) a finite number of modes are chosen and the 
eigenvalues are determined.  Using the modes corresponding to the reciprocal lattice vectors 
in the one mode approximation ($(m,n)=(\pm 1,0),(0,\pm 1), (1,-1), (-1,1)$ and 
the $(0,0)$ mode gives four eigenvectors that are always negative and thus irrelevant 
and three eigenvalues that have the form $-DQ^2$.  The smallest $D$ arises 
from $b_{0,0}$ mode and can be determined analytically if only this mode is used 
(the other eigenvalues correspond to $D \approx 3, 9$).  Since this is the smallest 
$D$ it determines the diffusion constant in the lattice.  The solution is 
\be
\label{eq:diffcnst}
D  = 3\psi^2+r+1+9 A_t^2/8.  
\ee
The accuracy of Eq. (\ref{eq:diffcnst}) was tested by 
numerically measuring the diffusion constant for the simulations shown in
Fig. (\ref{fig:diff1}).  In this calculation the envelope of profile of $\delta\psi$ was fit 
to a gaussian ($Ae^{-r^2/2\sigma^2}$) and the standard deviation ($\sigma$) was measured.  
The diffusion constant can be obtained by noting that the solution of a   
diffusion equation (i.e., $\partial C/\partial t = D \nabla^2 C$) is 
$C \propto e^{-r^2/4D t}$, i.e.,  $\sigma^2=D t/2$.  In Fig. (\ref{fig:diff3}) 
$\sigma^2$ is plotted as a function of time and the slope of this curve 
gives, $D \approx 1.22$.  This is quite close the value predicted by 
Eq. (\ref{eq:diffcnst}) which is $1.25$.

\section{Simple PFC Model: Applications}
\label{sec:app}

	In this section several application of the PFC model will be considered that 
highlight the flexibility of the model.  In Sec. (\ref{sec:grbo}) the energy of 
a grain boundary separating two grains of different orientation is considered.  
The results are compared with the Read-Shockley equation\cite{rs50} and shown 
to agree quite well for small orientational mismatch.  This calculation, in part,
provides evidence that the interaction between dislocations is correctly captured 
by the PFC model, since the grain boundary energy contains a term that is due to 
the elastic field set up by a line of dislocations.   In Sec. (\ref{sec:epi}) 
the technologically important process of liquid phase epitaxial growth is considered.
Numerical simulations are conducted as a function of mismatch strain and show 
how the model naturally produces the buckling instability and nucleation of dislocations. 
In Sec. (\ref{sec:hard}) the yield strength of poly- (nano-) crystalline materials is 
examined.  This is a phenomena that requires many of the features of the PFC model 
(i.e., multi orientations, elastic and plastic deformations, grain boundaries) that 
are difficult to incorporate in standard uniform phase field models.  The yield strength 
is examined as a function of grain size and the reverse Hall-Petch effect is observed.  
Finally some very preliminary numerical simulations are presented in Sec. (\ref{sec:other})
to demonstrate the versatility of the technique.  This section includes 
simulations of grain growth, crack propagation and reconstructive phase transitions.

\subsection{Grain Boundary Energy}
\label{sec:grbo}

\begin{figure}[btp]
\epsfxsize=8cm \epsfysize=8cm
\epsfbox{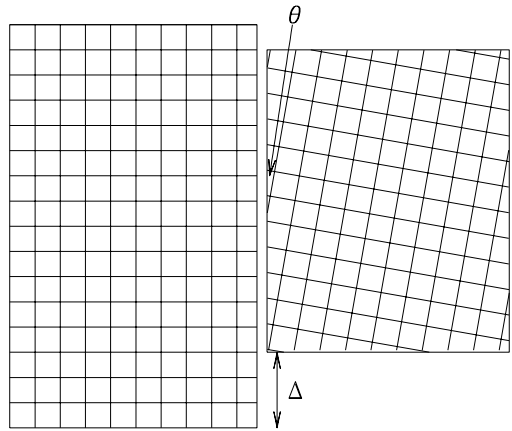}
\caption{Schematic of a grain boundary.}
\label{fig:RSzero}
\end{figure}

	The free energy density of a boundary between two grains that 
differ in orientation is largely controlled by geometry.  In a finite size 
two-dimensional system the parameters that control this energy are the orientational 
mismatch, $\theta$ and an offset distance $\Delta$, as shown in Fig. (\ref{fig:RSzero}).  
For small $\theta$, $\theta$ controls the number of dislocations 
per unit length and $\Delta$ controls the average core energy.  For an infinite grain boundary 
$\Delta$ becomes irrelevant, unless the distance between dislocation is an integer number 
of lattice constants (and the integer is relatively small).  Nevertheless 
it is straightforward to determine a lower bound on the grain boundary energy in the small 
$\theta$ limit, by directly relating the dislocation density to $\theta$ and assuming that 
the dislocation cores can always find the minimum energy location.  The later assumption 
restricts the calculation to providing a lower bound on the grain boundary energy. 

	For small $\theta$, Read and Shockley \cite{rs50} were able to 
derive an expression for the grain boundary energy, assuming the 
dislocation core energy was a constant independent of geometry.
In two dimensions the energy/length of the grain boundary is \cite{club},
\be
\frac{F}{L} = E_{core}+\frac{b^2Y_2}{8\pi d} \lt(1-\ln\lt(\frac{2\pi a}{d}\rt)\rt).
\ee
where $b$ is the magnitude of the Burger's vector, $a$ is the size of the dislocation 
core, $d$ is the distance between dislocations, $Y_2$ is the two dimensional Young's modulus 
and $E_{core}$ is the energy/length of the dislocation core.  To estimate the minimum core 
energy it is convenient to assume the core energy is proportional to the size of 
the core\cite{club}, i.e.,  $E_{core} = Ba^2$, where $B$ is an unknown constant.  
The total energy/length then becomes 
\be
\frac{F}{L} = Ba^2+\frac{b^2Y_2}{8\pi d} \lt(1-\ln\lt(\frac{2\pi a}{d}\rt)\rt).
\ee
To obtain a lower bound on $F/L$ the unknown parameter $B$ is chosen to 
minimize $F/L$, i.e., $B$ is chosen to satisfy, $d(F/L)/da = 0$, which 
gives $Ba^2 = b^2Y_2/16\pi d$.  Thus the free energy per unit length is,
\be
\label{eq:whatever}
\frac{F}{L} = \frac{b^2Y_2}{8\pi d} \lt(\frac{3}{2}-\ln\lt(\frac{2\pi a}{d}\rt)\rt).
\ee
Furthermore, from purely geometrical considerations, the distance between 
dislocations is $d=a/\tan(\theta)$, where $\theta$ is the orientational mismatch.  Finally
in the small angle limit ($\tan(\theta) \approx \theta$) Eq. (\ref{eq:whatever}) reduces to,
\be
\label{eq:rs}
\frac{F}{L} = \frac{bY_2}{8\pi} \theta \lt(\frac{3}{2}-\ln\lt(2\pi\theta\rt)\rt)
\ee
where the dislocation core size $b$ was assumed to be equal to the lattice constant $a$.

\begin{figure}[btp]
\epsfxsize=8cm \epsfysize=8cm
\epsfbox{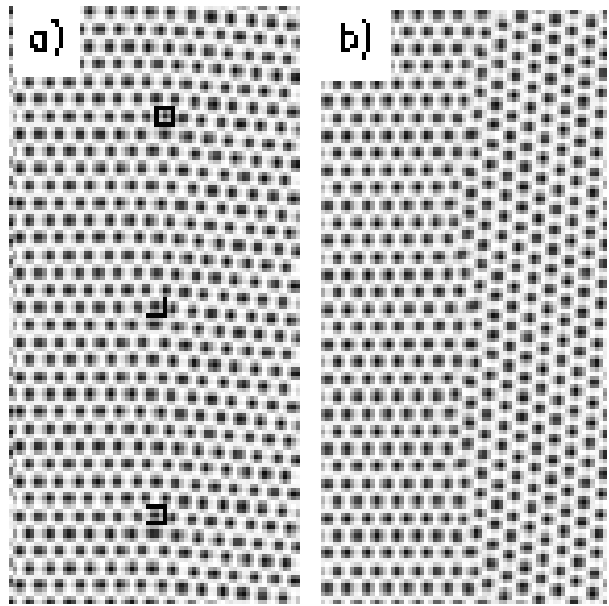}
\caption{In this figure the grey scale corresponds to the magnitude 
of the field $\psi$ for a grain boundary mismatch of $\theta=5.8^o$ 
and $\theta=34.2^o$ in figures a) and b) respectively.  In Fig. a) 
squares have been placed at defect sites.}
\label{fig:grbo}
\end{figure}

	To examine the validity of Eq. (\ref{eq:rs}) the grain boundary energy was measured as 
a function of angle.  In these simulations numerical Method I (see App. (\ref{app:metI})) 
was used with the parameter set $(r,\psi_o,\noise,\Delta x,\Delta t)=(-4/15,1/5,0,\pi/4,0.01)$.  
The initial condition was constructed as follows.  On a periodic grid of size 
$L_x \times L_y$, a triangular solution (i.e., Eq. (\ref{eq:tri})) for $\psi$ 
was constructed in one orientation between $0< x<L_x/4$ and $3L_x/4 < x < L_x$.  
In the center of the simulation (i.e., $L_x/4<x<3L_x/4$)
a triangular solution of a different orientation was constructed.   A small slab of 
supercooled liquid was placed between the two crystals so as not to influence the nature 
of the grain boundary that emerged.  The systems were then evolved for a time of $t=10,000$, 
after which the grain boundary energy was measured.  Small portions of sample configurations are 
shown in Fig. (\ref{fig:grbo}) for $\theta=5.8^o$ and $\theta=34.2^o$ (the grain boundary 
energy is symmetric around $30^o$). 
As expected the Read-Shockley description of a grain boundary is consistent with 
the small angle configuration.  In contrast the large angle grain boundary is much more 
complicated and harder to identify individual dislocations.

\begin{figure}[btp]
\epsfxsize=8cm \epsfysize=8cm
\epsfbox{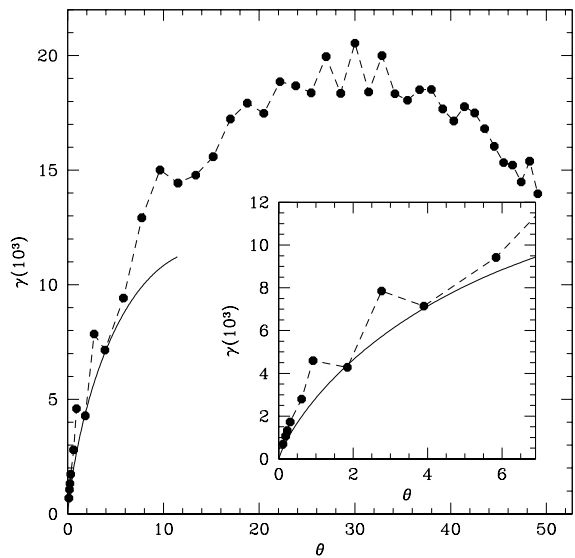}
\caption{In this figure the grain boundary energy is plotted 
as a function of mismatch orientation.  The points correspond 
to numerical simulations of the PFC model and the solid line 
corresponds to Eq. (\ref{eq:rs}).}
\label{fig:RSone}
\end{figure}

\begin{figure}[btp]
\epsfxsize=8cm \epsfysize=8cm
\epsfbox{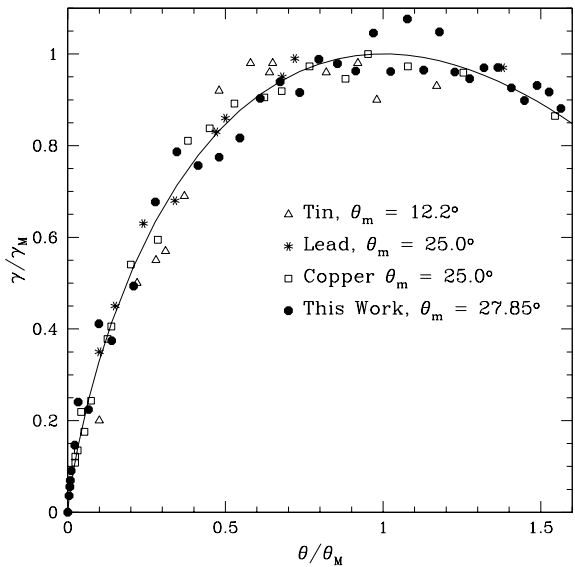}
\caption{In this figure the grain boundary energy of the PFC model is compared with 
experiments on Tin\cite{ac52}, Lead\cite{ac52} and Copper\cite{gr59}.}
\label{fig:RStwo}
\end{figure}

The measured grain boundary energy is compared with Eq. (\ref{eq:rs}) in Fig. (\ref{fig:RSone}).
As expected Eq. (\ref{eq:rs}) provides an adequate description for small angles but 
not for large angles.  The Read-Shockley equation does fit 
the measured result for all $\theta$ reasonably well if coefficients that enter the 
equation are adjusted, as has been observed in experiment \cite{ac52,gr59}.  
This fit is shown in Fig. (\ref{fig:RStwo}).

	The situation is obviously more complicated in three dimension since another 
degree of freedom exists.  This degree of freedom can be visualized by considering 
taking one of the crystals shown in Fig. (\ref{fig:RSzero}) and rotating it out of 
the page.  The extra degree of free can lead to interesting phenomena, such as 
coincident site lattices that significantly alter the grain boundary energy.  
The PFC model should provide an excellent tool for studying such phenomena since 
it is purely a geometrical effect that is naturally incorporated in the PFC approach.

\subsection{Liquid Phase Epitaxial Growth}
\label{sec:epi}

	Liquid phase epitaxial growth is a common industrial method used to grow 
thin films that are coherent with a substrate.  The properties of such films depend
on the structural integrity of the film.  Unfortunately flat defect-free heteroepitaxial 
films of appreciable thickness are often difficult to grow due to morphological 
instabilities induced by the anisotropic strain arising from the mismatch between 
film and substrate lattice constants.  Consequently, 
there has been a tremendous amount of scientific effort devoted to understanding the 
morphological stability of epitaxially grown films
\cite{at72,g82,s89,ys93,mg99,hmrg02,ekhg02,mb74,pb85,pnlc93,fkehr96,zc96,fn96,
rpmbs99,jpbh93,jptbbh93,ong93,gn99,ghf02,bjnrt96,osmyn98,rk91,ans92,gv96}.

    The stability and resulting structural properties of epitaxial
films are often compromised by at least two distinct processes which reduce
the anisotropic strain.  In one process, small mounds or ridges form as the
surface buckles or corrugates to reduce the overall strain in the film.
This instability to buckling can be predicted by considering the linear stability
of an anisotropically strained film as done by Asaro and Tiller \cite{at72} and
Grinfeld \cite{g82}. The initial length scale of the buckling is determined by a
competition between the reduction in overall elastic energy which prefers mounds
and surface tension and gravity both of which favor a flat interface.  
Another mechanism that reduces strain is the nucleation of misfit dislocations
which can occur when the energy of a dislocation loop is comparable
with the elastic energy of the strained film.  Matthews and Blakeslee \cite{mb74} 
and many others \cite{pb85,pnlc93,fkehr96,zc96,fn96,rpmbs99} have used various 
arguments to provide an expression for the critical height at which a flat 
epitaxially grown film will nucleate misfit dislocations.

    The two mechanisms are often considered separately but it is
clear that surface buckling can strongly influence the nucleation of
misfit dislocations.  Typically, as the film begins to grow,
it will deform coherently by the Asaro-Tiller-Grinfeld
instability.  This leads initially to a roughly sinusoidal film thickness with 
a periodicity close to the most unstable mode in a linear analysis.  As time 
increases, the sinusoidal pattern grows in amplitude and develops cusps
or local regions of high curvature \cite{jpbh93,jptbbh93,ong93,gn99} with
a periodicity similar to that of the initial instability although some
coarsening may occur \cite{mg99,ong93,gn99}.  Eventually, the stress at the 
cusps become too large and a periodic array of misfit dislocations appear which 
reduces the roughness of the film. These dislocations eventually climb to the 
film/substrate interface.  

\begin{figure}[btp]
\epsfxsize=8cm \epsfysize=8cm
\epsfbox{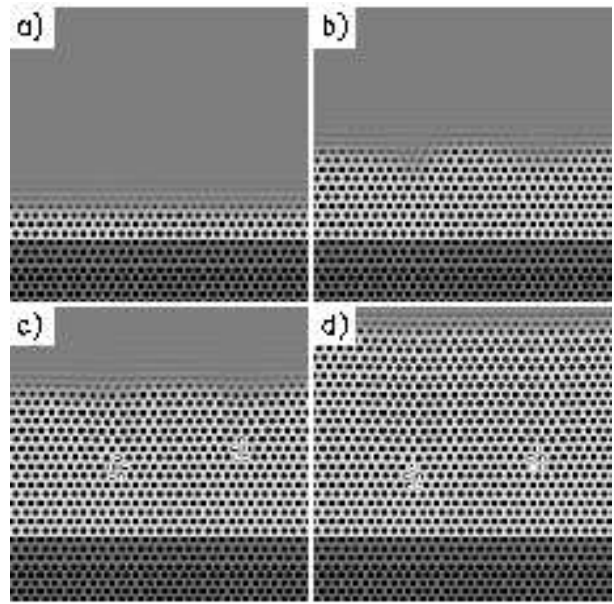}
\caption{Epitaxial growth.  Figures a), b), c) and d) correspond 
to times t=150, 300, 450 and 600 respectively.  The grey-scale is proportional 
to the local density (i.e., $\psi$) in the film and the liquid. 
The substrate is highlighted by a darker grey background.  To highlight  
nucleated dislocations, small white dots were placed on atoms near 
the two dislocation cores that appear in this configuration.}
\label{fig:aaa__70__small}
\end{figure}

	The purpose of this section is to illustrate how the PFC method can be exploited 
to examine surface buckling and dislocation nucleation in liquid phase epitaxial 
growth.  Modeling this process requires a slight modification of the model to 
incorporate a substrate that has a different lattice constant than the growing 
film.  This can be accomplished by changing the operator $\omega$ given 
in Eq. (\ref{eq:disp}) to be,
\be
\omega=r+(q^2+\nabla^2)^2,
\ee
where the parameter $q$ controls the lattice constant of the growing 
film and is set to one in the substrate.  To incorporate a constant mass 
flux the field $\psi$ was fixed to be, $\psi_{\ell}$ at a constant 
distance ($L=100\Delta x$) above the film.

\begin{figure}[btp]
\epsfxsize=8cm \epsfysize=7.7cm
\epsfbox{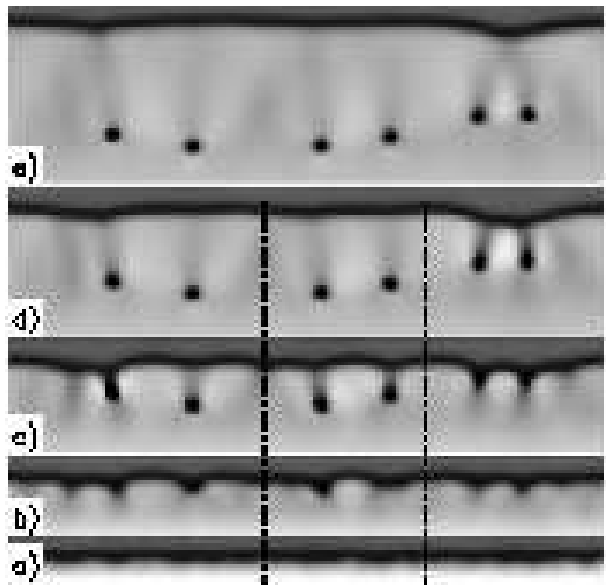}
\caption{Epitaxial growth.  Figures a), b), c), d) and e) correspond 
to times t=150, 300, 450, 600 and 750 respectively.  The grey-scale is proportional 
to the free energy density.  To highlight the excess strain energy in the film 
the grey scale near the defect was saturated.  The region enclosed by dashed lines 
corresponds to the configuration shown in Fig. (\ref{fig:aaa__70__small}).}
\label{fig:aaa70en}
\end{figure}

\begin{figure}[btp]
\epsfxsize=8cm \epsfysize=8cm
\epsfbox{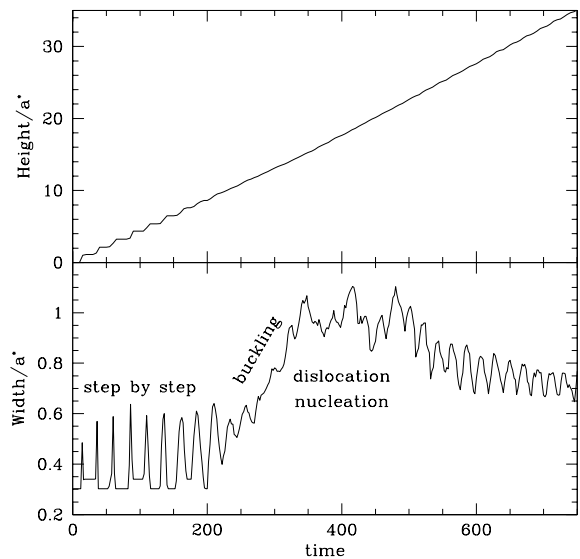}
\caption{Epitaxial growth.  In Figs. a) and b) the average 
film/liquid interface height and width is shown as a function of time. 
Both the width and height have been scaled by $a^*=2\pi$, which is 
the one mode approximation for the distance between layers in the 
appropriate direction}.  
\label{fig:hw}
\end{figure}

	Numerical simulations were conducted using Method I (see App. (\ref{app:metI}))
for the parameters $(r,\psi_{\ell},\Delta x,\Delta t)=(-1/4,.282,.785,0.0075)$.  The 
width of the film grown was $L_{x} = 8192 \Delta x$, corresponding 
to a width of roughly 900 particles.  The initial condition was such that 
eight layers of substrate atoms resided at the bottom of the 
simulation cell with a supercooled ($r=-1/4$,$\psi_{\ell}=.282$)  
liquid above it.  A small portion of a simulation is shown in 
Fig. (\ref{fig:aaa__70__small}), for $q=0.93$.  As can be seen in this figure, and 
in Fig. (\ref{fig:aaa70en}) the film initially grows in a uniform fashion before becoming 
unstable to a buckling or mounding instability.  The film then nucleates dislocations in 
the valleys where the stress is the largest.  After the dislocations nucleate the 
liquid/film interface grows in a more regular fashion.  To highlight the local elastic energy, 
the free energy is plotted in Fig. (\ref{fig:aaa70en}).  As can be seen in this 
figure, elastic energy builds up in the valleys during the buckling instability and is 
released when dislocations appear.  The behavior of the liquid/film interface was 
monitored by calculating the average interface height and width.  Both quantities are 
plotted in Fig. (\ref{fig:hw}).  The data shown in this figure is representative of all
simulations conducted at different mismatch strains, but the precise details 
varied from run to run.  In all cases the width initially 
fluctuates around $a^*/2$ (where $a^*$ is the thickness of a film layer) during the 
`step by step` growth.  The average width then increases during buckling and decreases 
when dislocations nucleate.  While these quantities are difficult to measure {\it in situ} 
there is experimental evidence for this behavior in SiGe/Si heterostructures \cite{ghf02}.

\begin{figure}[btp]
\epsfxsize=8cm \epsfysize=8cm
\epsfbox{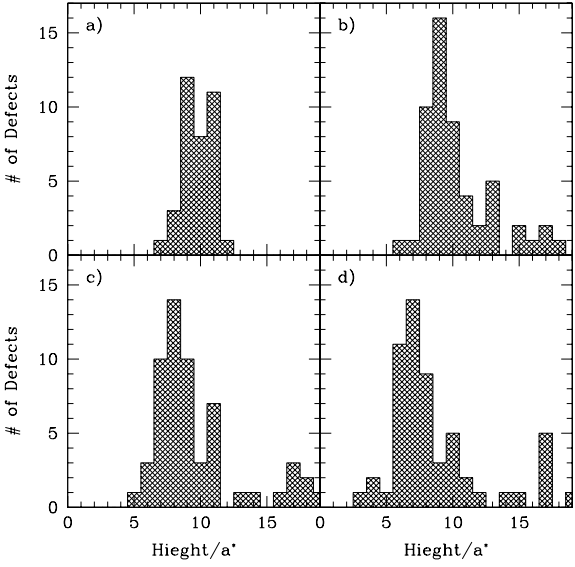}
\caption{Epitaxial growth.  In this figure a histogram of the number of 
defects is shown as a function of height above the substrate.  Figs. a), b), c) and d) 
correspond to $t=300, 450, 600$ and $1000$.}
\label{fig:runthist}
\end{figure}

	Assigning a value to the critical height, $H_c$ at which 
dislocations nucleate is very subjective.  Typically a first wave of 
dislocations is nucleated at a density that is determined by the 
buckling instability.  Since this is not the correct density to reduce 
the strain to zero a subsequent buckling and dislocation occurs above 
the first wave.  To complicate manners the nucleated dislocations climb 
towards the substrate/film interface.   To illustrate these points 
the dynamics of a sample distribution of defects is shown as function of height
in Fig. (\ref{fig:runthist}).  As can be seen in this figure the first wave 
of dislocations appears roughly between a film height of six and thirteen 
layers.  Comparison of Figs. (\ref{fig:runthist}b) and (\ref{fig:runthist}d) 
shows that as time evolves the overall distribution of dislocation climbs toward the 
surface.  To obtain an operational definition of $H_c$, the 
average height, $\bar{H}(t)$ of the first wave of dislocations was monitored as 
a function of time.  Typically $\bar{H}(t)$ is a maximum when all dislocations in 
the first wave have appeared and then decreases as the dislocation climb to the 
substrate/film interface.  $H_c$ was defined as the maximum value of $\bar{H}(t)$.

\begin{figure}[btp]
\epsfxsize=8cm \epsfysize=8cm
\epsfbox{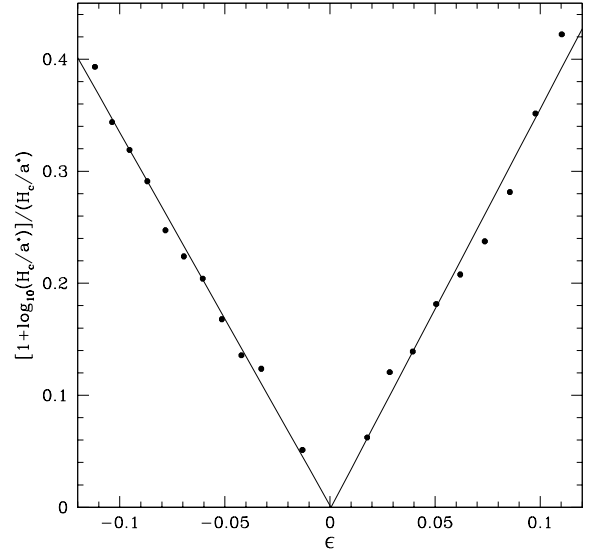}
\caption{Epitaxial growth.  In this figure the $H_c$ is the critical height 
as defined in the text and $\epsilon$ is the mismatch strain between the 
film and substrate.}
\label{fig:matblak}
\end{figure}

	The critical height, as defined in the preceding paragraph, was calculated 
as a function of mismatch strain, $\epsilon = (a_{film}-a_{substrate})/a_{substrate}$.  
The equilibrium lattice constant in the film $a_{film}$ was obtained by 
assuming it was directly proportional to $1/q$ (note, in the one mode approximation, 
$a=2\pi/(\sqrt(3)q/2)$) and determining the constant of proportionality by 
interpolating to where the critical height diverges.  The numerical data was 
compared  with the functional form proposed by 
Matthews and Blakeslee \cite{mb74} , i.e., 
\be
H_c \propto  \frac{1}{\epsilon}\left(1+\log_{10}\left[\frac{H_c}{a^*}\right]\right),
\ee  
in Fig. (\ref{fig:matblak}).  This comparison indicates that 
the data is consistent with 
a linear relationship between $\epsilon$ and $[1+\log(H_c/a^*)]/(H_c/a^*)$, where    
the constant of proportionality depends on whether a 
compressive or tensile load is applied to the substrate.

\subsection{Material Hardness}
\label{sec:hard}

	It is well known that mechanical properties of materials 
depend crucially on the microstructure and grain size.  For example,  
Hall and Petch \cite{h51,p53} calculated that for large grain sizes, 
the yield strength of a material is inversely proportional to the 
square of the average grain radius.  This result is due to the 
pileup of dislocations at grain boundaries and has been verified in 
many materials including Fe alloys \cite{cp55,hp57,jk90}, 
Ni \cite{hspja86}, Ni-P alloys \cite{lww90}, Cu \cite{crkg89} and 
Pd \cite{crkg89}.  However, for very small grain sizes the Hall-Petch 
relationship must break down, since the yield strength cannot diverge.   
Experimentally it is found that materials "soften" at very small grain 
sizes, such that the yield strength begins to decrease when the 
grain sizes become of the order of tens of nanometers.   This `inverse' Hall-Petch 
behavior has been observed in in Ni-P alloys \cite{lww90}, Cu and Pd \cite{crkg89} 
and molecular dynamics experiments \cite{sdj98,svdj99}.  
Determining the precise the crossover length scale and mechanisms  
of material breakdown has become increasingly important in technological   
processes as interest in nano-crystalline materials (and nano-technology 
in general) increases.  

	The purpose of this section is to demonstrate how the PFC approach 
can be used to study the influence of grain size on material strength.  
In these simulations a poly-crystalline sample was created by heterogenous 
nucleation (see Sec. (\ref{sec:graingrowth}) for details) in a system with 
periodic boundary conditions in both the $x$ and $y$ directions.  A small 
coexisting liquid boundary of width $200\Delta x$ was included on either 
side of the sample.  To apply a strain the particles near 
the liquid/crystal boundary (i.e., within a distance of $16\Delta x$) were `pulled` 
by coupling these particles to a moving field that fixed the particle positions.  Initially  
the system was equilibrated for a total time of $4000$ ($2000$ before the field was applied 
and $2000$ after).  An increasing strain was modeled by moving the field every so many 
time steps in such a manner that the size of the polycrystal increased by $2\Delta x$.  
To facilitate relaxation, $\psi$ was extrapolated to the new size after every movement of the 
external field.  The parameters of the simulations to follow were 
$(r,\psi_{sol},\psi_{liq},L_x,L_y,\Delta x,\Delta t)
=(-0.3,0.312,2048\Delta x,2048\Delta x,0.377,0.79,0.05)$ and the pseudo-spectral 
numerical method described in App. (\ref{app:metII}) was used.

\begin{figure}[btp]
\epsfxsize=8cm \epsfysize=8cm
\epsfbox{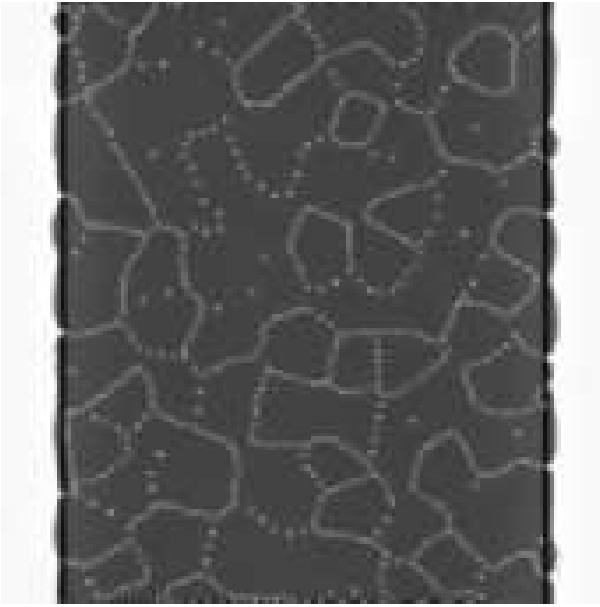}
\caption{In this figure the grey scale corresponds to the local 
energy density before a strain is applied.  The dark black regions on the left 
and right of the figure are the regions that are coupled to the external 
field.}
\label{fig:eab0}
\end{figure}

\begin{figure}[btp]
\epsfxsize=8cm \epsfysize=8cm
\epsfbox{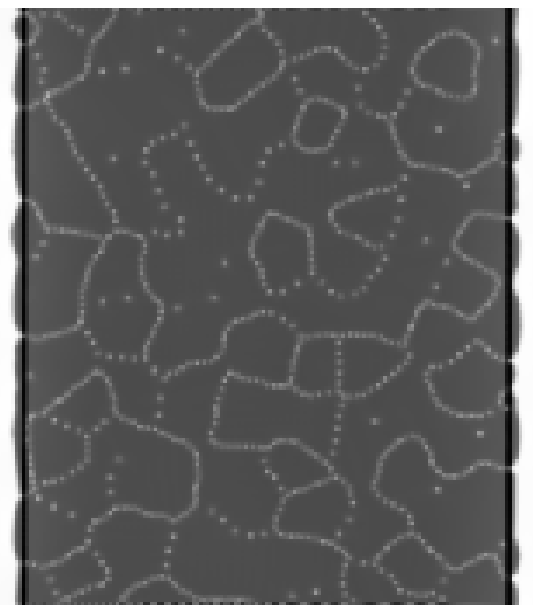}
\caption{This figure is the same as Fig. (\ref{fig:eab0}) except at a 
strain of $1.9\%$.}
\label{fig:eab32}
\end{figure}

\begin{figure}[btp]
\epsfxsize=8cm \epsfysize=8cm
\epsfbox{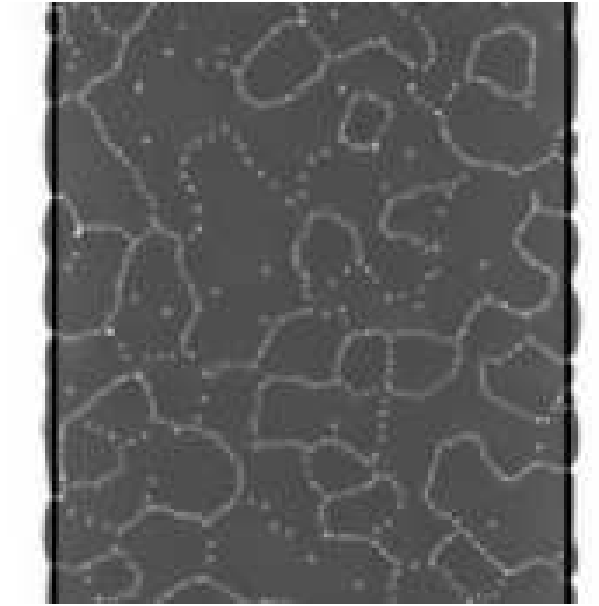}
\caption{This figure is the same as Fig. (\ref{fig:eab0}), except at 
at strain of $7.8\%$.}
\label{fig:eab128}
\end{figure}

	A sample initial configuration is shown in Fig. (\ref{fig:eab0}).  This particular 
sample contains approximately 100 grains with an average grain radius of 35 particles.  
As can be seen in this figure there exists a large variety (i.e., distribution 
of mismatch orientations) of grain boundaries as would exist in a realistic poly-crystalline 
sample.  The same configuration is shown after it has been stretched in the $x$ direction
in Figs. (\ref{fig:eab32}) and (\ref{fig:eab128}) corresponding to strains of 
$1.9\%$ and $7.8\%$ respectively.

\begin{figure}[btp]
\epsfxsize=8cm \epsfysize=8cm
\epsfbox{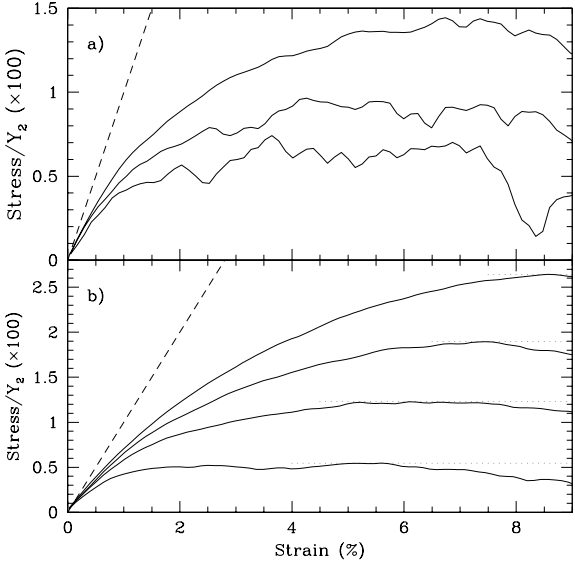}
\caption{In Fig. a) the stress is plotted as a function of strain for 
a system with an average grain radius of 
35 particles.  The solid lines from top to bottom in a) correspond to strain rates 
of $24 \times 10^{-6}$, $12 \times 10^{-6}$ and $6 \times 10^{-6}$ respectively.
In Fig. b) the solid lines from top to bottom correspond to systems with 
average grain sizes of $70$, $50$, $35$ and $18$ particles respectively.
In both a) and b) the dashed line corresponds to a unit slope.}
\label{fig:ststEbth}
\end{figure}

	As the poly-crystalline sample is pulled the total free energy 
was monitored and used to calculate the stress, i.e., ${\rm stress} \equiv  d^2 F/d\zeta^2$, 
where $\zeta$ is the relative change in the width of the crystal.  Stress-strain curves 
are shown in Fig. (\ref{fig:ststEbth}) as a function of grain size and strain rate.  
In all cases the stress is initially a linear of function of strain until plastic 
deformation occurs and the slope of the stress-strain curve decreases. 
In Fig. (\ref{fig:ststEbth}a) the influence of strain rate is examined for 
the initial configuration shown in Fig. (\ref{fig:eab0}).  It is clear from 
this figure that strain rate plays an strong role in determining the maximum 
stress that a sample can reach, or the yield stress, as has been observed in 
experiments \cite{sm60}.  The yield strength increases as the strain 
rate increases as would be expected.

\begin{figure}[btp]
\epsfxsize=8cm \epsfysize=8cm
\epsfbox{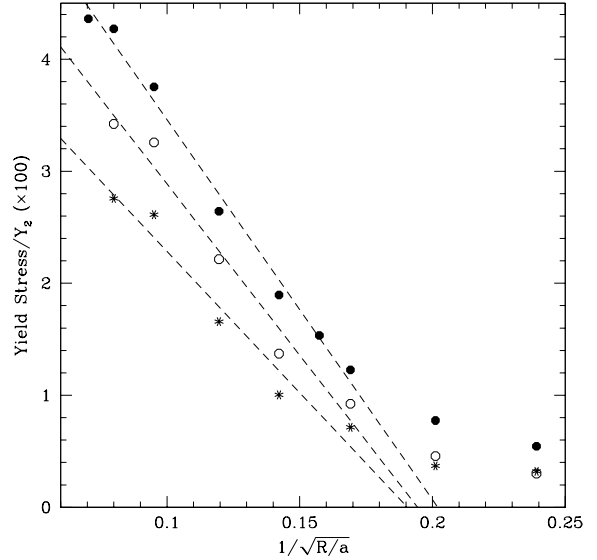}
\caption{In this figure the yield stress is plotted as function of 
average grain radius.
The solid, empty and starred points correspond to strain rates of 
of $24 \times 10^{-6}$, $12 \times 10^{-6}$ and $6 \times 10^{-6}$ respectively.
The dashed lines are guides to the eye.}
\label{fig:rHP}
\end{figure}

\begin{figure}[btp]
\epsfxsize=8cm \epsfysize=8cm
\epsfbox{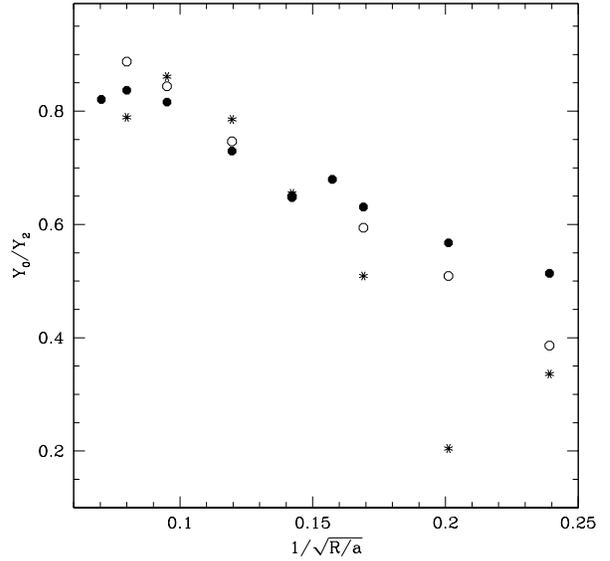}
\caption{The elastic moduli $Y_0$ (see text) is plotted as a 
function grain radius in this figure.
The solid, empty and starred points correspond to strain rates of 
of $24 \times 10^{-6}$, $12 \times 10^{-6}$ and $6 \times 10^{-6}$ respectively.}
\label{fig:elast}
\end{figure}

	The influence of grain size on the stress-strain relationship
is shown in Fig. (\ref{fig:ststEbth}b) for four grain sizes.  The initial 
slope of the stress-strain curve (which will be denoted $Y_0$ in what 
follows) increases with increasing grain size as does the maximum 
stress, or yield stress, sustained by the sample.    
The yield strength and elastic moduli ($Y_0$) are plotted as a function of inverse grain 
size in Figs. (\ref{fig:rHP}) and (\ref{fig:elast}) respectively for several strain rates.  
For each strain rate the 
yield stress is seen to be inversely proportional to the square root of the 
average grain size, however the constant of proportionality decreases with 
decreasing strain rate.  Thus the PFC approach is able to reproduce the 
inverse Hall-Petch effect or the softening of nano-crystalline materials. 

	It would be interesting to observe the cross-over to the normal 
Hall-Petch effect where the yield stress decreases with increasing grain 
size.  However, it is important to note that the initial conditions in 
these simulations was set up to explicitly remove the Hall-Petch mechanism, i.e., 
each nano-crystal was defect free.  In addition thermal fluctuations were 
not included in the simulations.  Nevertheless it is unclear whether or not a crossover 
may occur, due to the fact that low angle grain boundaries may act as sources 
of movable dislocations.   Further study of this interesting phenomena 
for larger grain sizes would be of great interest.

\subsection{Other Phenomena}
\label{sec:other}

	There are many phenomena that the PFC method can be used to 
explore.   To illustrate this a few small simulations were conducted to 
examine a number of interesting phenomena of current interest.  In 
the next few sections some preliminary results are shown for grain growth, 
crack propagation and reconstructive phase transitions.

\subsubsection{Grain Growth}
\label{sec:graingrowth}

	When a liquid is supercooled just below the melting temperature small 
crystallites can nucleate homogeneously or heterogeneously.  The crystallites 
will grow and impinge on neighboring crystallites forming grain boundaries. 
Depending on the temperature and average concentration the final state (i.e., 
in the infinite time limit) may be a single crystal or a coexistence of liquid 
and crystal phase since there exists a miscibility gap in density for some 
regions of the phase diagrams.  For deep temperatures quenches the liquid 
is unstable to the formation of a solid phase and initially an amorphous sample 
is created very rapidly which will evolve into a poly-crystalline sample and 
eventually become a single crystal (in the infinite time limit).  
All these phenomena can be studied with the simple PFC model considered in this paper.

\begin{figure}[btp]
\epsfbox{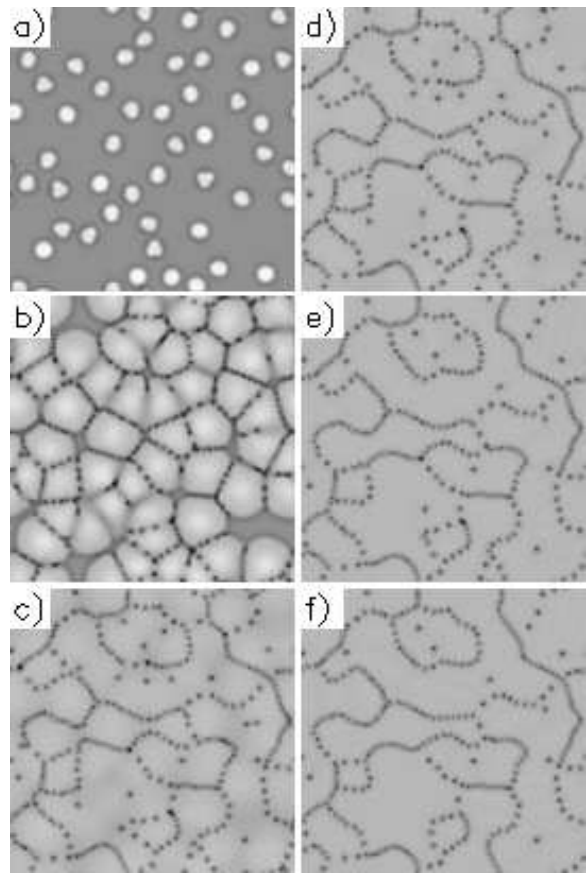}
\caption{Heterogenous nucleation and grain growth. In this figure the grey scale 
corresponds to the smoothed local free energy.  The figures a), b), c), d), e) 
and f) correspond to times 50, 200, 1000, 3000, 15000, 50000 respectively.}
\label{fig:graingrowth}
\end{figure}

	In this section the PFC model is used to examine the heterogenous nucleation 
of a poly-crystalline sample from a supercooled liquid state.  A simulation containing 
fifty initial seeds (or nucleation sites) was conducted.  The initial seeds were identical 
to those described in Sec. (\ref{sec:dynamics}) as were all other relevant parameters.
The results of the simulations are shown in Fig. (\ref{fig:graingrowth}).  
Comparison of Figs. (\ref{fig:graingrowth}b) and 
(\ref{fig:graingrowth}c) shows that there is a wide distribution of grain boundaries 
each with a different density of dislocations (which appear as black dots in 
the figure).  Comparison of (\ref{fig:graingrowth}c) with later configurations 
indicates that the low angle grain boundaries disappear much more rapidly than 
the large angle ones.  The simple reason is that it is easy for one or two 
dislocations to glide in such a manner as to reduce the overall energy (this is 
usually accompanied with some grain rotation).  The simulation was run for 
up to a time of $t=50,000$ (or approximately 1,200 diffusion times) and contained 
approximately 15,000 particles.  The simulation took roughly 70 hours of cpu on a 
single alpha chip processor (xp1000).

\subsubsection{Crack Propagation}
	
\begin{figure}[btp]
\epsfxsize=8cm \epsfysize=8cm
\epsfbox{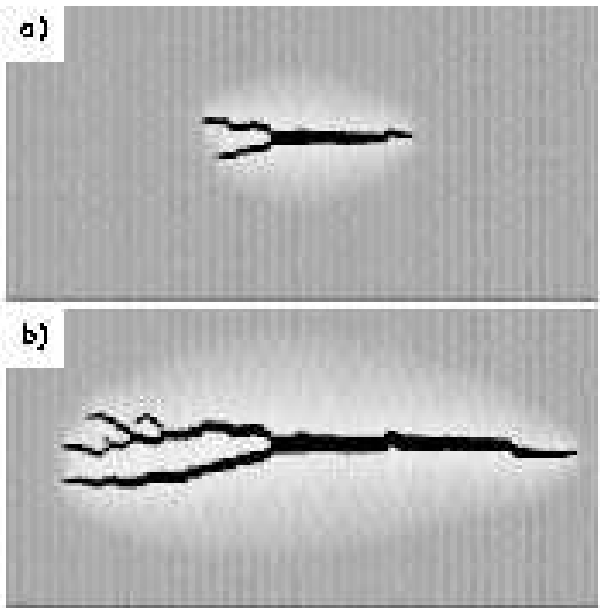}
\caption{In this figure a portion of a simulation is shown where the grey scale corresponds 
to the local energy density. The size of both figures is $2048 \Delta x \times 1024 \Delta x$,
where $\Delta x = \pi/3$. Figs. a) and b) are at times $t=25,000$ and $65,000$ after 
the rip was initiated respectively.}
\label{fig:rip}
\end{figure}

	The PFC model can be used to study the propagation of a crack in 
ductile (but not brittle) material.  To illustrate this phenomena a preliminary 
simulation was conducted on a periodic system of size $(4096\Delta x,1024\Delta x)$ 
for the parameters $(r,\psi_o,\Delta x,\Delta t)=(-1.0,0.49,\pi/3,.05)$. 
Initially a defect free crystal was set up in the simulation cell that had no 
strain in the $x$ direction and a $10\%$ strain in the $y$ direction.  A notch of size 
$20\Delta x \times 10 \Delta x$ was cut out of the center of the simulation cell and 
replaced with a coexisting liquid ($\psi=0.79$).  The notch provides a nucleating cite 
for a crack to start propagating.  A sample simulation is shown in Fig. (\ref{fig:rip}).

\subsubsection{Reconstructive Phase Transitions}

\begin{figure}[btp]
\epsfbox{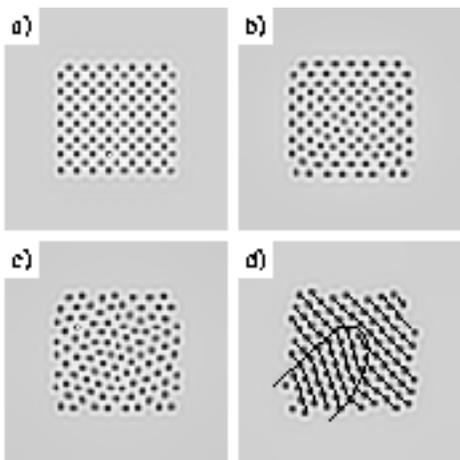}
\caption{In this figure the grey scale corresponds the field $\psi$. 
Figs. a), b), c) and d) correspond to times, $t=2,20,40$ and  $180$ 
respectively. In Fig. d) the solid lines are guides to the eye.}
\label{fig:recon}
\end{figure}

	The simple PFC model can be used to study a phase transition 
from a state with square symmetry to one with triangular symmetry.  
In the model described by Eq. (\ref{eq:dimfree}) a state with square symmetry 
is metastable, i.e., a state with square symmetry will remain unchanged unless 
boundary conduction or fluctuations are present.  Boundary conduction or 
fluctuations allow for the nucleation of a lower energy state which 
in this particular model is the state of triangular symmetry discussed 
in Sec. (\ref{sec:twodimpd}).  A small simulation was performed to illustrate 
this phenomena.  In this simulation a crystallite with square symmetry coexisting 
with a liquid was created as an initial condition.  The parameters for this simulation 
were $(r,\psi_{liq},\psi_{sol},\Delta x,\Delta t) = (1.0,0.68,0.52,1.0,.02)$.
The simulations depicted in Fig. (\ref{fig:recon}), show the spontaneous 
transition from square structure to the triangular structure.  Two 
variants of the triangular structure (differing by a rotation of 30 
degrees) form in the new phase as highlighted in 
Fig. (\ref{fig:recon}d).   

A better method for studying this phenomena is to create a free energy 
that contains both square and triangular symmetry equilibrium states.  
This can be done by including a $|\vec{\nabla}\psi|^4$ term (which favors 
square symmetry) in the free energy.  This is, unfortunately not the most 
convenient term for numerical simulations.   A better approach is 
to simply couple two fields in the appropriate manner as was done in 
an earlier publication \cite{ekhg02}.  In either case an initial poly-crystalline 
state can be created of one crystal symmetry.

\section{Summary}
\label{sec:sum}

	The purpose of this was paper was to introduce the PFC method of 
studying non-equilibrium phenomena involving elastic and plastic deformations and 
then to show how the technique can be applied to many phenomena. 
Those phenomena included epitaxial growth, material hardness, grain
growth, reconstructive phase transitions, crack propagation, and
spinodal decomposition.  In the future, we intend to extend this model to study these
phenomena in three dimensions.

\section{Acknowledgments}
This work was supported by a Research Corporation grant CC4787 (KRE),
NSF-DMR Grant 0076054 (KRE), the Natural Sciences and
Engineering Research Council of Canada (MG), and {\it le Fonds Qu\'eb\'ecois
de la Recherche sur la Nature et les Technologies\/} (MG).

\section{Appendix: Numerical Methods}
\label{app:nummet}

	Equation (\ref{eq:eom}) was numerically solved using the two 
different methods work as described below.  In what follows 
the subscripts $n$, $i$ and $j$ are integers that corresponds to 
the number of time steps and distance along the $x$ and $y$ directions 
of the the lattice respectively.  Time and space units are recovered 
by the simple relations, $t=n\Delta t$, $x=i\Delta x$ and $y=j\Delta x$.

\subsection{Method I}
\label{app:metI}

	In method I an Euler discretization scheme was used for the time 
derivative and the `spherical laplacian' approximation was used to calculate 
all Laplacians.  For this method the discrete dynamics read,
\be
\psi_{n+1,i,j} = \psi_{n,i,j}+\nabla^2 \mu_{n,i,j},
\ee
where $\mu_{n,i,j}$ is the chemical potential given by;
\be
\mu_{n,i,j} = (r+(1+\nabla^2)^2)\psi_{n,i,j}-\psi_{n,i,j}^3.
\ee
All Laplacians were evaluated as follows,
\ben
&&\nabla^2f_{n,i,j}=\Big(\big( 
f_{n,i+1,j}+f_{n,i-1,j}+f_{n,i,j+1}\nline
&+&f_{n,i,j-1}\big)/2+\big(f_{n,i+1,j+1}+f_{n,i-1,j+1}\nline
&+&f_{n,i-1,j+1}+f_{n,i-1,j+1}\big)/4-3f_{n,i,j}\Big)/(\Delta x)^2.
\een

\subsection{Method II}
\label{app:metII}

	In method II an Euler algorithm was again used for the time 
step, except that a simplifying assumption was made to evaluate $(r+(1+\nabla^2)^2)\psi_{n,i,j})$ 
in Fourier space.  In this approach the fourier transform of $\psi_{n,i,j}$ was numerically 
calculated then multiplied by $w(q)$ and then an inverse fourier transform was numerically 
evaluated to obtain an approximation to $(r+(1+\nabla^2)^2)\psi_{n,i,j})$.  If $w(q)$ 
is chosen to be $w(q)=r+(1-q^2)^2$ then, to within numerical accuracy, there is no 
approximation.  In this work $w(q)$ was chosen to be $r+(1-q^2)^2$ if $w(q) < -2.5$ and 
$w(q)=-2.5$ otherwise.  Thus $w(q)$ is identical to the exact result for wavevectors 
close to $q=1$, i.e., the wavelengths of interest.  The advantage of introducing a 
large wavevector cutoff is that the most numerically unstable modes arise from the largest 
negative values of $w(q)$.  This allows the use of much larger time steps.  Other than this 
approximation the method is identical to Method I.


\begin{thebibliography}{99}
\bibitem{periodic}  In many cases it was nessesary to numerically study a 
`perfect` crystal at various wavelengths or orientations in a lattice with 
periodic boundary conditions.   In order to avoid boundary effects the 
width (and height) of the simulation cell should be an integer number of 
wavelength along the $x$ ($y$) direction.  In all cases where this was 
necessary (i.e., elastic constants, grain boundary energies, etc.) the 
width and height were first adjusted independently to be as close as 
possible to satisfying this condition.  Finally the wavelengths (in 
both $x$ and $y$ directions) and orientations were then adjusted to 
provide an exact match.
\end{thebibliography}
\end{document}